\documentclass[10pt,conference]{IEEEtran}

\IEEEoverridecommandlockouts

\let\origfigure\figure
\let\endorigfigure\endfigure

\renewenvironment{figure}[1][tbph]{%
    \origfigure[#1]%
    \centering
}{%
    \endorigfigure
}

\usepackage[utf8]{inputenc}
\usepackage{amsmath,amssymb,amsfonts,amsthm}
\usepackage{tikz}
\usepackage{float}
\usepackage{pgfplots}
\usetikzlibrary{shapes.geometric,arrows,positioning}
\usetikzlibrary{quantikz}
\usetikzlibrary{external}
\usetikzlibrary{zx-calculus}
\zxConfigureExternalSystemCallAuto

\usepackage{braket}
\usepackage[caption=false]{subfig}

\usepackage{caption}
\usepackage{graphicx}
\usepackage{mwe}
\usepackage{widetext}

\newtheorem{definition}{Definition}

\usepackage[boxed, algoruled]{algorithm2e}
\usepackage{array}
\usepackage{lipsum}
\usepackage{enumitem}
\usepackage{nicematrix}

\usepackage{todonotes}\reversemarginpar

\usepackage{multirow,bigdelim,dcolumn,booktabs}
\usepackage{graphicx}
\usepackage{mathtools}  

\usepackage[export]{adjustbox}
\usepackage{tabularx}

\usepackage{algorithm2e}

\adjustboxset*{center}


\tikzset{
diagonal fill/.style 2 args={fill=#2, path picture={
\fill[#1, sharp corners] (path picture bounding box.south west) -|
                         (path picture bounding box.north east) -- cycle;}},
reversed diagonal fill/.style 2 args={fill=#2, path picture={
\fill[#1, sharp corners] (path picture bounding box.north west) |- 
                         (path picture bounding box.south east) -- cycle;}}
}

\tikzset{
pauliY/.style={
anchor=center,
minimum height=0.5em,
minimum width=0.5em,
draw,
line width=\zxDefaultLineWidth,
diagonal fill={colorZxX}{colorZxZ}
}
}

\tikzset{
pauliX/.style={
anchor=center,
minimum height=0.5em,
minimum width=0.5em,
draw,
line width=\zxDefaultLineWidth,
fill=colorZxX
}
}

\tikzset{
pauliZ/.style={
anchor=center,
minimum height=0.5em,
minimum width=0.5em,
draw,
line width=\zxDefaultLineWidth,
fill=colorZxZ
}
}

\tikzset{
pauliPhase/.style={
anchor=center,
minimum height=1em,
minimum width=1em,
draw,
line width=\zxDefaultLineWidth,
fill=white
}
}

\tikzset{
redPlus/.style={
minimum height=0.5em,
minimum width=0.5em,
line width=0,
circle,
inner sep=-0.15em,
outer sep=-0.15em,
fill=colorZxX,
line width=0,
draw
}
}
\usepackage{filemod}

\begin{document}

\title{A recursively partitioned approach to architecture-aware ZX Polynomial synthesis and optimization}
\author{\IEEEauthorblockN{1\textsuperscript{st} David Winderl}
\IEEEauthorblockA{\textit{CIT Department of Computer Science} \\
\textit{Technical University of Munich}\\
 Boltzmannstra{\ss}e 3\\
Garching, Germany \\
david.winderl@tum.de}
\and
\IEEEauthorblockN{1\textsuperscript{st} Qunsheng Huang}
\IEEEauthorblockA{\textit{CIT Department of Computer Science} \\
\textit{Technical University of Munich}\\
Boltzmannstra{\ss}e 3\\
Garching, Germany \\
keefe.huang@tum.de}
\and
\IEEEauthorblockN{2\textsuperscript{rd} Christian B.~Mendl}
\IEEEauthorblockA{\textit{CIT Department of Computer Science and IAS}\\
\textit{Technical University of Munich}\\
Boltzmannstra{\ss}e 3\\
Garching, Germany \\
christian.mendl@tum.de}
}
\date{\today}
\maketitle
\begin{abstract}
The synthesis of quantum circuits from phase gadgets in the ZX-calculus facilitates quantum circuit optimization. Our work provides an alternative formulation for the architecture-aware synthesis algorithm of PauliOpt~\cite{PauliOpt} by replacing the stochastic approach of PauliOpt with a heuristic based search and utilizes a divide and conquer method to synthesize an optimized circuit from a ZX polynomial. We provide a comparison of our algorithm with PauliOpt and other state-of-the-art optimization libraries. While we note poorer performance for highly structured circuits, as in the QAOA formulation for Max-Cut, we demonstrate a significant advantage for randomized circuits, which highlights the advantages of utilizing an architecture-aware methodology.
\end{abstract}
\section{Introduction}
The noisy intermediate-scale quantum (NISQ) era in quantum computing is defined by noisy qubits with limited coherence times and restricted connectivity \cite{Kissinger2019}. 
Due to the short viability of each qubit, there is strong motivation to reduce the depth of a given circuit.
Indeed, this remains a popular research field that has motivated many circuit optimization strategies~\cite{Nam2018, Duncan2019, Amy_2014, Sivarajah_2020}.
However, the restricted connectivity of qubits complicates the direct optimization of a circuit as interactions between non-neighboring qubits are implemented by chains of two-qubit gates (classically SWAP gates) specified by the underlying hardware.
This problem, known as the routing problem \cite{Cowtan2019}, has gathered significant attention in recent years as routing gates often introduces many two-qubit gates.
Hence, most state-of-the-art circuit optimization, transpilation, or synthesis tools, such as BQSKit \cite{osti_1785933}, or tket \cite{Sivarajah_2020}, inherently include a routing algorithm that runs iteratively with optimization strategies.
In recent years, the ZX-calculus, a graphical language for the reasoning of linear maps, has emerged~\cite{Wetering_ZXCalc}. This language provides, from our point of view, the capability to formulate variational ans\"{a}tze~\footnote{Algorithms, consisting of an interplay between a classical optimizer and a quantum circuit encoding a mostly NP-Hard optimization problem.} in a more natural sense, by so-called ZX Polynomials.
In this realm, Griend et al.~\cite{GriendPP} introduced a \emph{architecture-aware} optimization strategy that performed circuit optimization while obeying the restricted connectivity of the underlying hardware, utilizing Z polynomials in the diagrammatic ZX-calculus representation.
They produced circuits that did not require an additional routing step but were limited to circuits that only utilized Z spiders.
Gogioso et al.~\cite{PauliOpt} provide an algorithm for synthesizing ZX polynomials based on a simulated annealing approach in the PauliOpt library. 
The efficient non-deterministic procedure to propagate CNOTs through a ZX polynomial allows a quick reduction in CNOT count without an exhaustive search of all possible synthesizable CNOTs.
One major advantage of this technique is the symmetric structure generated when constructing parity maps to the left and right of a ZX polynomial by extracting CNOTs, which is significant for quantum algorithms relying on repeated layers or structures as the series of CNOTs between layers cancel.

\medskip

The present work utilizes Gogioso et al.'s symmetrical ansatz and incorporates ideas from Griend et al., with two major extensions. First, we propose a novel heuristic-based search as opposed to the previous stochastic implementation in PauliOpt. Second, we provide a recursive divide-and-conquer formulation that improves optimization results. 
We compare the performance of this strategy with the prior implementation and compare results wtih existing state-of-the-art optimization routines, namely pyzx ~\cite{Kissinger2021PYZX} and tket~\cite{Sivarajah_2020}.
While the new method does not outperform other quantum circuit transpiliation algorithms on structured circuits, we demonstrate significant improvement in the context of random circuits.
In addition, we can identify possible weaknesses in prior approaches.

\section{Preliminaries}
\subsection{ZX-calculus}\label{sec:2_prel_zx_calculus}
The ZX-calculus is a rigorous graphical language representing arbitrary linear maps.
Initially introduced in \cite{Coecke2011}, the diagrammatic calculus was further extended in many works, such as \cite{Duncan2009, Jeandel2017, Vilmart2019}, and was shown to be complete for the $\frac{\pi}{4}$ Clifford fragment, which corresponds to the Clifford+T gate set \cite{Backens2014, Backens2015}. 
Given that the $\frac{\pi}{4}$ fragment is approximately universal \cite{Jeandel2018}, one can then approximate any arbitrary quantum circuit in terms of the ZX-calculus.
Converting from an arbitrary quantum circuit to the ZX-calculus representation is well understood and detailed in \cite{Cowtan2020,Coecke2017,Coecke2011}.

The main benefit of using the ZX-calculus representation (which we refer to as a ZX diagram) is the ease of simplification.
Supported by many rewrite rules detailing suitable substitutions \cite{Wetering_ZXCalc}, several quantum circuit optimization algorithms already employ ZX-calculus with comparable performance to state-of-the-art optimization algorithms or heuristics \cite{Kissinger2021PYZX, Cowtan2020,Duncan2019}. 

\subsection{Phase gadgets}\label{sec:phase_gadgets}
\begin{figure}[ht!]
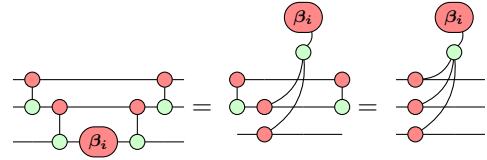

    \centering
    \begin{equation*}
    \begin{ZX}[math baseline=t1]
        &\\
        &\\
    \zxNone{} \rar & \zxX[a=c1]{}   &              &                    &                    & \zxX[a=c4]{} \rar    & \zxNone{}\\
    \zxNone{} \rar & \zxZ[a=t1]{}   & \zxX[a=c2]{} &                    & \zxX[a=c3]{}       & \zxZ[a=t4]{} \rar    &\zxNone{}\\
    \zxNone{} \rar & \zxNone{} \rar & \zxZ[a=t2]{} & \zxX[a=g]{\beta_i} & \zxZ[a=t3]{} \rar  &  \zxNone{} \rar      &\zxNone{}\\
    \ar[from=c1,to=t1]
    \ar[from=c1,to=c4]
    \ar[from=t1,to=c2]
    \ar[from=c2,to=c3]
    \ar[from=c2,to=t2]
    \ar[from=t2,to=g]
    \ar[from=g,to=t3]
    \ar[from=t3,to=c3]
    \ar[from=c3,to=t4]
    \ar[from=t4,to=c4]
    \end{ZX}
    =
    \begin{ZX}[math baseline=l1]
                &         & \zxX[a=pgp1]{\beta_i}             & \zxNone{}\\
                &         & \zxZ[a=pg1]{} \ar[u, bend right]  & \zxNone{}\\
        \zxX[a=c1]{} \rar & \zxNone{} \rar    & \zxNone{} \rar            & \zxX[a=c2]{}\\
        \zxZ[a=t1]{} \rar & \zxX[a=l1]{}\ar[ruu, bend right] \rar  & \zxNone{} \rar            & \zxZ[a=t2]{}\\
        \zxNone{} \rar    & \zxX{}\ar[ruuu, bend right] \rar  & \zxNone{} \rar            & \zxNone{}\\
    \ar[from=c1,to=t1]      
    \ar[from=c2,to=t2]
    \end{ZX}
    =
    \begin{ZX}[math baseline=l1]
                &                   & \zxX[a=pgp1]{\beta_i}             & \zxNone{}\\
                &                   & \zxZ[a=pg1]{} \ar[u, bend right]  & \zxNone{}\\
        \zxNone{} \rar  & \zxX{} \ar[ru, bend right] \rar       & \zxNone{} \rar  & \zxNone{}\\
        \zxNone{} \rar  & \zxX[a=l1]{}\ar[ruu, bend right] \rar & \zxNone{} \rar  & \zxNone{}\\
        \zxNone{} \rar  & \zxX{}\ar[ruuu, bend right] \rar      & \zxNone{} \rar  & \zxNone{}\\
    \end{ZX}
\end{equation*} 
    \caption{Classial (left) and compact (right) representation of an $X$ phase gadget in the ZX-calculus.}
    \label{fig:x_phase_gadget_zx}
\end{figure}

In this work, we utilize the \textit{phase gadget} formulation \cite{Cowtan2020} as part of the ansatz for optimization.
\begin{definition}\label{def:z_phase_gadet}
Z phase gadgets $\Phi_{(z,q)}(\alpha, n): \mathbb{C} \rightarrow \mathbb{C}, \; q,n \in \mathbb{Z^+}, \; q \ge n$ are a family of unitary maps with the recursive definition:
\begin{align*}
\Phi_{(z,q)}(\alpha, n) &= \text{CNOT}(q-n, q-n+1) \\
                        & \quad \cdot \big(I \otimes \Phi_{(p,q)}(\alpha, n-1) \big) \\
                        & \quad \cdot \text{CNOT}(q-n, q-n+1), \\
\Phi_{(z,q)}(\alpha, 0) &= R_z(\alpha)_q,
\end{align*}
where $\text{CNOT}(a, b)$, $a, b \in \mathbb{Z^+}$, is a CNOT-gate with control on qubit $a$ and target on qubit $b$ and $R_z$ is defined as follows:
$$
R_z(\alpha) = \begin{pmatrix}
	e^{-i \frac{\alpha}{2}} & 0 \\
	0 & e^{i \frac{\alpha}{2}}
\end{pmatrix}
$$. 
\end{definition}
\begin{definition}\label{def:x_phase_gadet}
$X$ phase gadgets $\Phi_{(x,q)}(\alpha, n): \mathbb{C} \rightarrow \mathbb{C}, \; q,n \in \mathbb{Z^+}, \; q \ge n$ are a family of unitary maps with the recursive definition:
\begin{align*}
\Phi_{(x,q)}(\alpha, n) &= \text{CNOT}(q-n+1, q-n) \\
                        & \quad \cdot \big(I \otimes \Phi_{(p,q)}(\alpha, n-1) \big) \\
                        & \quad \cdot \text{CNOT}(q-n+1, q-n), \\
\Phi_{(x,q)}(\alpha, 0) &= R_z(\alpha)_q,
\end{align*}
where $R_x$ is defined as follows:
$$
R_x(\alpha) = \begin{pmatrix}
	\cos (\frac{\alpha}{2}) & -i \sin (\frac{\alpha}{2}) \\
	-i \sin (\frac{\alpha}{2}) & \cos (\frac{\alpha}{2})
\end{pmatrix}
$$
\end{definition}
By definition, a phase gadget consists of a CNOT (reverse-CNOT) ladder, followed by a single Z (X) rotation and then a reversed CNOT (reverse-CNOT) ladder. However, we can express this construction in a more compact notation, utilizing rewrite rules in \cite{Cowtan2020}. An example $X$ phase gadget in both representations is shown in Fig.~\ref{fig:x_phase_gadget_zx}.
Additionally, one can append or remove the legs of the phase gadget by ``pulling out'' CNOTs (reverse-CNOTs) to the left and right.
We refer to a series of concatenated phase gadgets as a \textit{phase polynomial}, referring to phase polynomials with only Z phase gadgets as a Z phase polynomial and so forth. Please note that the natural extension towards phase polynomials are so-called Pauli polynomials, which allow non-pure assignment of the Pauli $\{X, Y, Z, I\}$ matrices on their legs. 
In this work, we additionally allow a linear transformation in $GF(2)$\footnote{$GF(2)$, the Galois field, describes a finite field of two elements (usually $0$ and $1$)} in phase polynomials, referred to as a global parity or parity map, for greater generality.
These are defined in Section.~\ref{sec:parity_maps}.

\subsection{Rewrite rules for ZX polynomials}\label{sec:rewrite_rules}
Two main rewrite rules are used for manipulating ZX polynomials in this work: CNOT gate propagation and commutation rules.

\subsubsection{Propagation of CNOT gates}
One can propagate a CNOT through an Z phase or X phase gadget by Eq.~\eqref{eq:propagation_cnot_z} and \eqref{eq:propagation_cnot_x} \cite{Cowtan2020}. Legs are added or removed depending on how the CNOT is oriented.
\begin{equation}\label{eq:propagation_cnot_z}
    \begin{aligned}
        \begin{ZX}
            \zxNone{}      &                            &                   & \zxZ[a=pgp1]{\alpha_i}            & \zxNone{}           &\zxNone{}\\
            \zxNone{}      &                    &                           & \zxX[a=pg1]{} \ar[u, bend right]  & \zxNone{}           &\zxNone{}\\
            \zxNone{} \rar & \zxX[a=c1]{}\ar[d] \rar        & \zxZ{} \ar[ru, bend right] \rar   & \zxNone{} \rar            & \zxX[a=c2]{} \ar[d] \rar   &\zxNone{}\\
            \zxNone{} \rar & \zxZ[a=t1]{} \rar              & \zxNone{} \rar            & \zxNone{} \rar            & \zxZ[a=t2]{} \rar   &\zxNone{}\\
                     & \zxNone{}                & \vdots{}              &                   &           & \\
            \zxNone{} \rar & \zxNone{} \rar             & \zxZ{}\ar[ruuuu, bend right] \rar & \zxNone{} \rar            & \zxNone{} \rar      &\zxNone{}\\
        \end{ZX}
        =
        \begin{ZX}
            \zxNone{}      &                            & \zxZ[a=pgp1]{\alpha_i}\\
            \zxNone{}      &                    & \zxX[a=pg1]{} \ar[u, bend right]\\
            \zxNone{} \rar & \zxZ{} \ar[ru, bend right] \rar    & \zxNone{}\\
            \zxNone{} \rar & \zxZ{} \ar[ruu, bend right] \rar   & \zxNone{}\\
             \zxNone{}  & \vdots{}              &       \\
            \zxNone{} \rar  & \zxZ{}\ar[ruuuu, bend right] \rar & \zxNone{}\\
        \end{ZX}
        &
        \begin{ZX}[math baseline=t1]
            \zxNone{}      &                    &                   & \zxZ[a=pgp1]{\alpha_i}            & \zxNone{}\\
            \zxNone{}      &            &                           & \zxX[a=pg1]{} \ar[u, bend right]  & \zxNone{}\\
            \zxNone{} \rar & \zxX[a=c1]{}\ar[d] \rar& \zxNone{} \rar            & \zxNone{} \rar            & \zxNone{}\\
            \zxNone{} \rar & \zxZ[a=t1]{} \rar      & \zxZ{} \ar[ruu, bend right] \rar  & \zxNone{} \rar            & \zxNone{}\\
                     & \zxNone{}        & \vdots{}              &                   &      \\
            \zxNone{} \rar & \zxNone{} \rar     & \zxZ{}\ar[ruuuu, bend right] \rar & \zxNone{} \rar            & \zxNone{}\\
        \end{ZX}
        =
        \begin{ZX}[math baseline=t1]
                    &                   & \zxZ[a=pgp1]{\alpha_i}            & \zxNone{}           &\zxNone{}\\
                    &                           & \zxX[a=pg1]{} \ar[u, bend right]  & \zxNone{}           &\zxNone{}\\
            \zxNone{} \rar  & \zxNone{} \rar            & \zxNone{} \rar            & \zxX[a=c1]{}\ar[d] \rar &\zxNone{}\\
            \zxNone{} \rar  & \zxZ{} \ar[ruu, bend right] \rar  & \zxNone{} \rar            & \zxZ[a=t1]{} \rar     &\zxNone{}\\
            \zxNone{}   & \vdots{}              &                   &           & \\
            \zxNone{} \rar  & \zxZ{}\ar[ruuuu, bend right] \rar & \zxNone{} \rar            & \zxNone{} \rar      &\zxNone{}\\
        \end{ZX}
    \end{aligned}
\end{equation}
\begin{equation}\label{eq:propagation_cnot_x}
    \begin{aligned}
        \begin{ZX}[math baseline=t1]
            \zxNone{}      &                    &                   & \zxX[a=pgp1]{\beta_i}         & \zxNone{}           &\zxNone{}\\
            \zxNone{}      &            &                           & \zxZ[a=pg1]{} \ar[u, bend right]  & \zxNone{}           &\zxNone{}\\
            \zxNone{} \rar & \zxX[a=c1]{}\ar[d] \rar& \zxNone[a=l1]{} \rar          & \zxNone{} \rar            & \zxX[a=c2]{} \ar[d] \rar   &\zxNone{}\\
            \zxNone{} \rar & \zxZ[a=t1]{} \rar      & \zxX{} \ar[ruu, bend right] \rar  & \zxNone{} \rar            & \zxZ[a=t2]{} \rar   &\zxNone{}\\
                     & \zxNone{}        & \vdots{}              &                   &           & \\
            \zxNone{} \rar & \zxNone{} \rar     & \zxX{}\ar[ruuuu, bend right] \rar & \zxNone{} \rar            & \zxNone{} \rar      &\zxNone{}\\
        \end{ZX}
        =
        \begin{ZX}[math baseline=l1]
                        &                   & \zxX[a=pgp1]{\beta_i} \\
                         &                              & \zxZ[a=pg1]{} \ar[u, bend right] \\
             \zxNone{} \rar     & \zxX{} \ar[ru, bend right] \rar   & \zxNone{} \\
             \zxNone{} \rar     & \zxX[a=l1]{} \ar[ruu, bend right] \rar    & \zxNone{} \\
              \zxNone{}     & \vdots{}              &       \\
             \zxNone{} \rar     & \zxX{}\ar[ruuuu, bend right] \rar & \zxNone{} \\
        \end{ZX}
        &
        \begin{ZX}[math baseline=l1]
            \zxNone{}      &                    &                   & \zxX[a=pgp1]{\beta_i}         & \zxNone{}\\
            \zxNone{}      &            &                           & \zxZ[a=pg1]{} \ar[u, bend right]  & \zxNone{}\\
            \zxNone{} \rar & \zxX[a=c1]{}\ar[d] \rar& \zxX{} \ar[ru, bend right] \rar   & \zxNone{} \rar            & \zxNone{}\\
            \zxNone{} \rar & \zxZ[a=l1]{} \rar      & \zxNone{} \rar            & \zxNone{} \rar            & \zxNone{}\\
                     & \zxNone{}        & \vdots{}              &                   &    \\
            \zxNone{} \rar & \zxNone{} \rar     & \zxX{}\ar[ruuuu, bend right] \rar & \zxNone{} \rar            & \zxNone{}\\
        \end{ZX}
        =
        \begin{ZX}[math baseline=l1]
                    &                   & \zxX[a=pgp1]{\beta_i}         & \zxNone{}           &\zxNone{}\\
                    &                           & \zxZ[a=pg1]{} \ar[u, bend right]  & \zxNone{}           &\zxNone{}\\
            \zxNone{} \rar  & \zxX{} \ar[ru, bend right] \rar   & \zxNone{} \rar            & \zxX[a=c1]{}\ar[d] \rar &\zxNone{}\\
            \zxNone{} \rar  & \zxNone[a=l1]{} \rar  & \zxNone{} \rar            & \zxZ[a=t1]{} \rar     &\zxNone{}\\
            \zxNone{}   & \vdots{}              &                   &           & \\
            \zxNone{} \rar  & \zxX{}\ar[ruuuu, bend right] \rar & \zxNone{} \rar            & \zxNone{} \rar      &\zxNone{}\\
        \end{ZX}
    \end{aligned}
\end{equation}

\subsubsection{Commutation rules for ZX phase gadgets}
In general, two ZX phase gadgets do not commute, and there is, to our knowledge, no general decomposition that swaps the order of two arbitrary ZX phase gadgets.
As an edge case, two ZX phase gadgets commute if and only if they are either of the same type (for example, two Z polynomials) or have an even number of legs on the same wires \cite{Yeung2020}.
\begin{equation}\label{eq:zx_equal}
    \begin{ZX}[math baseline=l1]
        &                   & \zxZ[a=pgp1]{\alpha_i}                & \zxNone{}&
        &                   & \zxX[a=pgp1]{\beta_i}             & \zxNone{}\\
        &                           & \zxX[a=pg1]{} \ar[u, bend right]  & \zxNone{}&
        &                           & \zxZ[a=pg1]{} \ar[u, bend right]  & \zxNone{}\\
    \zxNone{} \rar  & \zxZ{} \ar[ru, bend right] \rar   & \zxNone{} \rar            & \zxNone{}\rar&
    \zxNone{} \rar  & \zxX{} \ar[ru, bend right] \rar   & \zxNone{} \rar            & \zxNone{}\\
    \zxNone{} \rar  & \zxNone{} \rar            & \zxNone{} \rar            & \zxNone{}\rar&    
    \zxNone{} \rar  & \zxNone[a=l1]{} \rar          & \zxNone{} \rar            & \zxNone{}\\
    \zxNone{} \rar  & \zxZ{}\ar[ruuu, bend right] \rar  & \zxNone{} \rar            & \zxNone{}\rar&
    \zxNone{} \rar  & \zxX{}\ar[ruuu, bend right] \rar  & \zxNone{} \rar            & \zxNone{}\\
    \end{ZX}
    =
    \begin{ZX}[math baseline=l1]
        &                   & \zxX[a=pgp1]{\alpha_i}                & \zxNone{}&
        &                   & \zxZ[a=pgp1]{\beta_i}             & \zxNone{}\\
        &                           & \zxZ[a=pg1]{} \ar[u, bend right]  & \zxNone{}&
        &                           & \zxX[a=pg1]{} \ar[u, bend right]  & \zxNone{}\\
    \zxNone{} \rar  & \zxX{} \ar[ru, bend right] \rar   & \zxNone{} \rar            & \zxNone{}\rar&
    \zxNone{} \rar  & \zxZ{} \ar[ru, bend right] \rar   & \zxNone{} \rar            & \zxNone{}\\
    \zxNone{} \rar  & \zxNone{} \rar            & \zxNone{} \rar            & \zxNone{}\rar&    
    \zxNone{} \rar  & \zxNone[a=l1]{} \rar          & \zxNone{} \rar            & \zxNone{}\\
    \zxNone{} \rar  & \zxX{}\ar[ruuu, bend right] \rar  & \zxNone{} \rar            & \zxNone{}\rar&
    \zxNone{} \rar  & \zxZ{}\ar[ruuu, bend right] \rar  & \zxNone{} \rar            & \zxNone{}\\
    \end{ZX}
\end{equation}
\begin{equation}\label{eq:zx_not_equal}
    \begin{ZX}[math baseline=l1]
                &                   & \zxZ[a=pgp1]{\alpha_i}                & \zxNone{}&
                &                   & \zxX[a=pgp1]{\beta_i}             & \zxNone{}\\
                &                           & \zxX[a=pg1]{} \ar[u, bend right]  & \zxNone{}&
                &                           & \zxZ[a=pg1]{} \ar[u, bend right]  & \zxNone{}\\
        \zxNone{} \rar  & \zxZ{} \ar[ru, bend right] \rar   & \zxNone{} \rar            & \zxNone{}\rar&
        \zxNone{} \rar  & \zxX{} \ar[ru, bend right] \rar   & \zxNone{} \rar            & \zxNone{}\\
        \zxNone{} \rar  & \zxZ{}\ar[ruu, bend right] \rar   & \zxNone{} \rar            & \zxNone{}\rar&    
        \zxNone{} \rar  & \zxX[a=l1]{}\ar[ruu, bend right] \rar     & \zxNone{} \rar            & \zxNone{}\\
        \zxNone{} \rar  & \zxZ{}\ar[ruuu, bend right] \rar  & \zxNone{} \rar            & \zxNone{}\rar&
        \zxNone{} \rar  & \zxX{}\ar[ruuu, bend right] \rar  & \zxNone{} \rar            & \zxNone{}\\
        \end{ZX}
        \neq
        \begin{ZX}[math baseline=l1]
                &                   & \zxX[a=pgp1]{\alpha_i}                & \zxNone{}&
                &                   & \zxZ[a=pgp1]{\beta_i}             & \zxNone{}\\
                &                           & \zxZ[a=pg1]{} \ar[u, bend right]  & \zxNone{}&
                &                           & \zxX[a=pg1]{} \ar[u, bend right]  & \zxNone{}\\
        \zxNone{} \rar  & \zxX{} \ar[ru, bend right] \rar   & \zxNone{} \rar            & \zxNone{}\rar&
        \zxNone{} \rar  & \zxZ{} \ar[ru, bend right] \rar   & \zxNone{} \rar            & \zxNone{}\\
        \zxNone{} \rar  & \zxX{}\ar[ruu, bend right] \rar   & \zxNone{} \rar            & \zxNone{}\rar&    
        \zxNone{} \rar  & \zxZ[a=l1]{}\ar[ruu, bend right] \rar     & \zxNone{} \rar            & \zxNone{}\\
        \zxNone{} \rar  & \zxX{}\ar[ruuu, bend right] \rar  & \zxNone{} \rar            & \zxNone{}\rar&
        \zxNone{} \rar  & \zxZ{}\ar[ruuu, bend right] \rar  & \zxNone{} \rar            & \zxNone{}\\
    \end{ZX}
\end{equation}
Another exceptional case was identified during our experimentation, called the \textit{$\pi$-commutation rule}. Assume two phase-gadgets; if one of the phase gadgets has phase $\pi$ and both do not share an even number of legs, they commute if the phase of the second phase gadget is flipped, as shown in Eqs.~\eqref{eq:phase_commutation_pi_x} and~\eqref{eq:phase_commutation_pi_z}.
\begin{equation}\label{eq:phase_commutation_pi_x}
    \begin{ZX}
        \zxNone{}      &                            & \zxX[a=pgp1]{\beta_i}&
        \zxNone{}      &                            & \zxZ[a=pgp1]{\pi}\\
        \zxNone{}      &                    & \zxZ[a=pg1]{} \ar[u, bend right]&
        \zxNone{}      &                    & \zxX[a=pg1]{} \ar[u, bend right]\\
        \zxNone{} \rar & \zxNone{} \rar     & \zxNone{}\rar&
        \zxNone{} \rar & \zxZ{} \ar[ru, bend right] \rar    & \zxNone{}\\
        \zxNone{} \rar & \zxX{} \ar[ruu, bend right] \rar   & \zxNone{}\rar&
        \zxNone{} \rar & \zxZ{} \ar[ruu, bend right] \rar   & \zxNone{}\\
         \zxNone{}  & \vdots{}              &   &
         \zxNone{}  & \vdots{}              &   \\
         \zxNone{} \rar & \zxX{}\ar[ruuuu, bend right] \rar & \zxNone{}\rar&
        \zxNone{} \rar  & \zxZ{}\ar[ruuuu, bend right] \rar & \zxNone{}\\
    \end{ZX}
    =
    \begin{ZX}
        \zxNone{}      &                            & \zxZ[a=pgp1]{\pi}&
        \zxNone{}      &                            & \zxX[a=pgp1]{-\beta_i}\\
        \zxNone{}      &                    & \zxX[a=pg1]{} \ar[u, bend right]&
        \zxNone{}      &                    & \zxZ[a=pg1]{} \ar[u, bend right]\\
        \zxNone{} \rar & \zxZ{} \ar[ru, bend right] \rar    & \zxNone{}\rar&
        \zxNone{} \rar & \zxNone{} \rar     & \zxNone{}\\
        \zxNone{} \rar & \zxZ{} \ar[ruu, bend right] \rar   & \zxNone{}\rar&
        \zxNone{} \rar & \zxX{} \ar[ruu, bend right] \rar   & \zxNone{}\\
         \zxNone{}  & \vdots{}              &   &
         \zxNone{}  & \vdots{}              &   \\
         \zxNone{} \rar & \zxZ{}\ar[ruuuu, bend right] \rar & \zxNone{}\rar&
        \zxNone{} \rar  & \zxX{}\ar[ruuuu, bend right] \rar & \zxNone{}\\
    \end{ZX}
\end{equation}
\begin{equation}\label{eq:phase_commutation_pi_z}
    \begin{ZX}
        \zxNone{}      &                            & \zxX[a=pgp1]{\beta_i}&
        \zxNone{}      &                            & \zxZ[a=pgp1]{\pi}\\
        \zxNone{}      &                    & \zxZ[a=pg1]{} \ar[u, bend right]&
        \zxNone{}      &                    & \zxX[a=pg1]{} \ar[u, bend right]\\
        \zxNone{} \rar & \zxNone{} \rar     & \zxNone{}\rar&
        \zxNone{} \rar & \zxZ{} \ar[ru, bend right] \rar    & \zxNone{}\\
        \zxNone{} \rar & \zxX{} \ar[ruu, bend right] \rar   & \zxNone{}\rar&
        \zxNone{} \rar & \zxZ{} \ar[ruu, bend right] \rar   & \zxNone{}\\
         \zxNone{}  & \vdots{}              &   &
         \zxNone{}  & \vdots{}              &   \\
         \zxNone{} \rar & \zxX{}\ar[ruuuu, bend right] \rar & \zxNone{}\rar&
        \zxNone{} \rar  & \zxZ{}\ar[ruuuu, bend right] \rar & \zxNone{}\\
    \end{ZX}
    =
    \begin{ZX}
        \zxNone{}      &                            & \zxZ[a=pgp1]{\pi}&
        \zxNone{}      &                            & \zxX[a=pgp1]{-\beta_i}\\
        \zxNone{}      &                    & \zxX[a=pg1]{} \ar[u, bend right]&
        \zxNone{}      &                    & \zxZ[a=pg1]{} \ar[u, bend right]\\
        \zxNone{} \rar & \zxZ{} \ar[ru, bend right] \rar    & \zxNone{}\rar&
        \zxNone{} \rar & \zxNone{} \rar     & \zxNone{}\\
        \zxNone{} \rar & \zxZ{} \ar[ruu, bend right] \rar   & \zxNone{}\rar&
        \zxNone{} \rar & \zxX{} \ar[ruu, bend right] \rar   & \zxNone{}\\
         \zxNone{}  & \vdots{}              &   &
         \zxNone{}  & \vdots{}              &   \\
         \zxNone{} \rar & \zxZ{}\ar[ruuuu, bend right] \rar & \zxNone{}\rar&
        \zxNone{} \rar  & \zxX{}\ar[ruuuu, bend right] \rar & \zxNone{}\\
    \end{ZX}
\end{equation}
A detailed proof of this rule utilizing ZX-calculus rewrite rules is shown in Appendix~\ref{sec:proof_pi_commutation}.

\subsection{Parity maps}\label{sec:parity_maps}
Circuits described in the Clifford+T gate set can be decomposed into single-qubit rotation gates and CNOTs.
Our work describes rotation operations using a series of ZX phase gadgets, detailed in Section~\ref{sec:phase_gadgets}.
A parity map is a binary matrix representing the action of a series of CNOT operations.
We can construct the parity map of a series of CNOTs as follows: Starting with an identity matrix $I \in \mathbb{B}^{n \times n}$ for an $n$-qubit system, where each row represents a qubit. 
We then ``apply'' each CNOT operation to this matrix by elementary row addition modulo $2$ of the control qubit's row to that of the target qubit.
Then a subsequent Gaussian elimination solution of such a matrix using only row addition operations recovers a series of CNOTs that implements the action of the original series of CNOTs.
As the Gaussian elimination solution of a given parity map is not unique, one may find a more optimal series of CNOTs performing the same action. By restricting the chosen row operations, Kissinger et al.~\cite{Kissinger2019} provide an algorithm that synthesizes a architecture-aware CNOT circuit from parity maps by utilizing Steiner-trees, named the \emph{recursive Steiner-Gauss}.
While a runtime analysis of this algorithm was not provided, it was concurrently developed by Nash et al.~\cite{Nash_2020}; the algorithm runs in $\mathcal{O}(d (|V| + |E|))$, where $d$ is the number of terminals in the connectivity graph of the underlying architecture, and $E$ and $V$ are the number of edges and vertices in the graph. In the worst case, by assuming an all-to-all connectivity we can assume: $|E| = q^2$, $|V| = q$ and $d = \mathcal{O}(q)$. Hence the runtime of the \emph{recursive Steiner-Gauss} is upper bounded by $\mathcal{O}(q^3)$.

%
%
\section{Methods}
Our proposed algorithm accepts a circuit in phase-polynomial representation and returns an optimized architecture-aware quantum circuit.
We divide our proposed algorithm into two parts: \emph{Simplification} and \emph{Synthesis}. 
\emph{Simplification} describes the process of reducing the number of gadgets inside of a ZX Polynomial, and \emph{Synthesis} describes the process of extracting CNOTs from a ZX polynomial with the intent of reducing total CNOT count.

\subsection{Simplification}
Our goal is to reduce the number of Z and X phase gadgets in a given ZX polynomial.

We develop a strategy based on peephole optimization, inspired by Nam et al.~\cite{Nam2018}. The general goal of the simplification strategy is to merge and remove as many phase gadgets as possible. We need to consider the following three operations:
\begin{itemize}
\item \textbf{Moving phase gadgets through the circuit:}\\
We apply commutation rules outlined in Section~\ref{sec:rewrite_rules}, shifting phase gadgets through the circuit
\item \textbf{Removing phase gadgets:}\\
We can remove a phase gadget if and only if its phase corresponds to $0$ or $2\pi$. In this case, the Z or X spider in the middle will be the identity matrix, and the phase gadget ``collapses''.
\item \textbf{Merging phase gadgets:}\\
We can merge two-phase gadgets if and only if all of their legs are equal by the following rule derived by Cowtan et al.~\cite{Cowtan2020}:
\end{itemize}
\begin{align}\label{eq:phase_merging_zx}
    \begin{ZX}
        \zxNone{}      &                    & \zxX[a=pgp1]{\beta_i}&
        \zxNone{}      &                    & \zxX[a=pgp1]{\beta_j}\\
        \zxNone{}      &                    & \zxZ[a=pg1]{} \ar[u, bend right]&
        \zxNone{}      &                    & \zxZ[a=pg1]{} \ar[u, bend right]\\
        \zxNone{} \rar & \zxX{} \ar[ru, bend right] \rar    & \zxNone{}\rar&
        \zxNone{} \rar & \zxX{} \ar[ru, bend right] \rar    & \zxNone{}\\
        \zxNone{} \rar & \zxX{} \ar[ruu, bend right] \rar   & \zxNone{}\rar&
        \zxNone{} \rar & \zxX{} \ar[ruu, bend right] \rar   & \zxNone{}\\
        \zxNone{}      & \vdots{}              &   &
        \zxNone{}      & \vdots{}              &   \\
        \zxNone{} \rar & \zxX{}\ar[ruuuu, bend right] \rar & \zxNone{}\rar&
        \zxNone{} \rar & \zxX{}\ar[ruuuu, bend right] \rar & \zxNone{}\\
    \end{ZX}
    =
    \begin{ZX}
        \zxNone{}      &                            & \zxX[a=pgp1]{\beta_i + \beta_j}\\
        \zxNone{}      &                    & \zxZ[a=pg1]{} \ar[u, bend right]\\
        \zxNone{} \rar & \zxX{} \ar[ru, bend right] \rar    & \zxNone{}\\
        \zxNone{} \rar & \zxX{} \ar[ruu, bend right] \rar   & \zxNone{}\\
        \zxNone{}      & \vdots{}              &   \\
        \zxNone{} \rar & \zxX{}\ar[ruuuu, bend right] \rar & \zxNone{}\\
    \end{ZX}
    \\
    \begin{ZX}
        \zxNone{}      &                            & \zxZ[a=pgp1]{\alpha_i}&
        \zxNone{}      &                            & \zxZ[a=pgp1]{\alpha_j}\\
        \zxNone{}      &                    & \zxX[a=pg1]{} \ar[u, bend right]&
        \zxNone{}      &                    & \zxX[a=pg1]{} \ar[u, bend right]\\
        \zxNone{} \rar & \zxZ{} \ar[ru, bend right] \rar    & \zxNone{}\rar&
        \zxNone{} \rar & \zxZ{} \ar[ru, bend right] \rar    & \zxNone{}\\
        \zxNone{} \rar & \zxZ{} \ar[ruu, bend right] \rar   & \zxNone{}\rar&
        \zxNone{} \rar & \zxZ{} \ar[ruu, bend right] \rar   & \zxNone{}\\
        \zxNone{}      & \vdots{}              &   &
        \zxNone{}      & \vdots{}              &   \\
        \zxNone{} \rar & \zxZ{}\ar[ruuuu, bend right] \rar & \zxNone{}\rar&
        \zxNone{} \rar & \zxZ{}\ar[ruuuu, bend right] \rar & \zxNone{}\\
    \end{ZX}
    =
    \begin{ZX}
        \zxNone{}      &                            & \zxZ[a=pgp1]{\alpha_i + \alpha_j}\\
        \zxNone{}      &                    & \zxX[a=pg1]{} \ar[u, bend right]\\
        \zxNone{} \rar & \zxZ{} \ar[ru, bend right] \rar    & \zxNone{}\\
        \zxNone{} \rar & \zxZ{} \ar[ruu, bend right] \rar   & \zxNone{}\\
        \zxNone{}      & \vdots{}              &   \\
        \zxNone{} \rar & \zxZ{}\ar[ruuuu, bend right] \rar & \zxNone{}\\
    \end{ZX}
\end{align}
As a simple example, we examine the two Hadamards, which cancel themselves:
\begin{equation}\label{eq:two_hadamarads}
    \begin{quantikz}
        \lstick{}& \gate{H}     & \gate{H}      & \qw \\
    \end{quantikz}
    =
    \begin{quantikz}
        \lstick{}& \ghost{H}\qw      & \qw      & \qw \\
    \end{quantikz}
\end{equation}
We can use Wetering et al.~\cite{Wetering_ZXCalc} to decompose the two Hadamard gates into a set of $Z$- and $X$-rotations, which we can rewrite into the following ZX Polynomial:
\begin{equation*}
    \begin{ZX}[math baseline=t1]
                &                   & \zxZ[a=pgp1]{\frac{\pi}{2}}                   & \zxNone{}&
                &                   & \zxX[a=pgp1]{\frac{\pi}{2}}                   & \zxNone{}&
                &                   & \zxZ[a=pgp1]{\frac{\pi}{2}}               & \zxNone{}&
                &                   & \zxZ[a=pgp1]{\frac{\pi}{2}}                   & \zxNone{}&
                &                   & \zxX[a=pgp1]{\frac{\pi}{2}}               & \zxNone{}&
                &                   & \zxZ[a=pgp1]{\frac{\pi}{2}}                   & \zxNone{}\\
                &                           & \zxX[a=pg1]{} \ar[u, bend right]      & \zxNone{}&
                &                           & \zxZ[a=pg1]{} \ar[u, bend right]      & \zxNone{}&
                &                           & \zxX[a=pg1]{} \ar[u, bend right]      & \zxNone{}&
                &                           & \zxX[a=pg1]{} \ar[u, bend right]      & \zxNone{}&
                &                           & \zxZ[a=pg1]{} \ar[u, bend right]      & \zxNone{}&
                &                           & \zxX[a=pg1]{} \ar[u, bend right]      & \zxNone{}\\
        \zxNone{} \rar  & \zxZ{}\ar[ru, bend right] \rar    & \zxNone{} \rar                & \zxNone{}\rar&
        \zxNone{} \rar  & \zxX{}\ar[ru, bend right] \rar    & \zxNone{} \rar                & \zxNone{}\rar&
        \zxNone{} \rar  & \zxZ[a=t1]{}\ar[ru, bend right] \rar  & \zxNone{} \rar                & \zxNone{}\rar&
        \zxNone{} \rar  & \zxZ{}\ar[ru, bend right] \rar    & \zxNone{} \rar                & \zxNone{}\rar&
        \zxNone{} \rar  & \zxX{}\ar[ru, bend right] \rar    & \zxNone{} \rar                & \zxNone{}\rar&
        \zxNone{} \rar  & \zxZ{}\ar[ru, bend right] \rar    & \zxNone{} \rar                & \zxNone{}\\
    \end{ZX}
\end{equation*}
We utilize a repeated \emph{merge} and \emph{remove} operation to repeatedly remove pairs of matching phase gadgets, which also returns the identity.

Formulated as an algorithm, as a first step, we commute each phase gadget through the ZX polynomial when possible. If we find another phase gadget with the same number of legs, we combine the two phase gadgets. Otherwise, we will restore the position of the phase gadget we moved through the circuit. As a second step, we remove all phase gadgets with a phase of $0$ or $2\pi$. We then continue those two steps until convergence.
Note that the PauliOpt library provides a similar simplification procedure.
Nevertheless, since it was developed congruently in a previous university project (see \cite{Winderl2022ZXPolynomial}) and not described in their publication, we still decided to include a description of the process in this work.
\subsection{Synthesis}
\begin{figure}[ht]
    \centering
    \begin{quantikz}[row sep=0.5em]
        \lstick{} & \qw    & \gate[5, nwires=3]{P_l} & \qw    & \gate[5, nwires=3]{\text{ZX}} & \qw    & \gate[5, nwires=3]{P_r} & \qw    &\qw &\\
        \lstick{} & \qw    &                         & \qw    &                               & \qw    &                         & \qw    &\qw &\\
        \lstick{} & \vdots &                         & \vdots &                               & \vdots &                         & \vdots &    &\\
        \lstick{} & \qw    &                         & \qw    &                               & \qw    &                         & \qw    &\qw &\\
        \lstick{} & \qw    &                         & \qw    &                               & \qw    &                         & \qw    &\qw &\\
    \end{quantikz}
    \caption{Splitting of a general ZX polynomial into regions.}\label{fig:zx_region_split}
\end{figure}
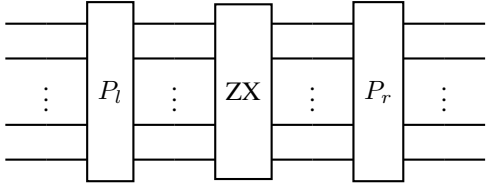

Given a ZX polynomial, we can remove a leg from a phase gadget and propagate the resulting pair of CNOTs using the commutation rules in Section~\ref{sec:rewrite_rules} to the left and right of the ZX polynomial. 
For a more complex circuit, we could amass many CNOTs to the left and right of the circuit and apply the recursive Steiner-Gauss algorithm to the resulting parity map, which we refer to as a \emph{parity region}, for further optimization.
This results in three regions for optimization: a ZX region and two parity regions to its left ($P_l$) and right ($P_r$), as shown in Figure~\ref{fig:zx_region_split}.
In a ZX polynomial, local structures might be difficult to resolve when propagating CNOTs to the outer parts of the polynomial. For this reason, we also optimize regions \emph{in between} the ZX polynomial. 
Therefore, we created two processes for synthesizing a ZX polynomial: \emph{optimization} of CNOTs and \emph{splitting} of the ZX polynomial.

\subsubsection{Optimization}
In the optimization step, the goal is to remove as many CNOTs as possible.
We determine the cost of removing a leg and propagating the resulting CNOTs out of the circuit as the change (increase or decrease) of CNOTs is required by the final output circuit.
We denote this as $e(c, t)$, where $c$ and $t$ refer to the control and target qubits of the propagated CNOT.
\begin{equation}\label{eq:effect_total}
    e(c, t) = \underbrace{e(ZX, c, t)}_{\textnormal{cost on ZX polynomial}} + \; \underbrace{e(P_l, c, t) + e(P_r, c, t)}_{\textnormal{cost on parity regions}}.
\end{equation}
There are two contributions to the cost: the change in the ZX polynomial $e(ZX, c, t)$ and the change in the parity regions $e(P_l, c, t),\;e(P_r, c, t)$.
For determining the change in the ZX polynomial, we adopt the approach by  Gogioso et al.~\cite{PauliOpt}, where Kruskals algorithm is used to find a minimal spanning tree representing the connectivity of the underlying architecture.
This then produces a minimal CNOT placement that is architecture specific. 
The complexity of the algorithm is polynomially bounded with $\mathcal{O}(q^2)$ for a single phase gadget, with an upper bound of $\mathcal{O}(nq^2)$ for a ZX polynomial with $n$ gadgets acting on $q$ qubits.
Additional steps demonstrated by Gogioso et al., such as caching CNOT costs, further reduce time to solution but are not considered when determining algorithmic complexity~\cite{PauliOpt}.
Determining the change in the parity regions is more computationally expensive.
Here, we compare the number of steps required by the recursive-Steiner-Gauss algorithm before and after the parity regions are modified; this equivalently determines the change in produced CNOTs.
We denote this effect in Eq.~\eqref{eq:effect_steiner_gauss} given an initial parity region $P$, a method that finds the number of steps of the Steiner-Gauss of a given parity region $c(P)$ and a modified parity region $P'$.
\begin{equation}\label{eq:effect_steiner_gauss}
    e(P, c, t) = c(P) - c(P').
\end{equation}
Then, given a  function $\texttt{propagate}$, which modifies the ZX and parity regions by propagating a CNOT with control $c$ and target $t$, we iterate over the possible combination of controls and targets to compute $e(P_l, c, t)$, $e(P_r, c, t)$ and $e(\text{ZX}, c, t)$ and greedily update the overall ZX polynomial whenever there is an improvement (see Algorithm~\ref{alg:gauss_optimization}).
%
\begin{algorithm}
    \caption{Optimization using the Gaussian elimination algorithm for optimization}\label{alg:gauss_optimization}
    \SetKw{KwIn}{in}
    \KwData{current $P_l$, $P_r$, $\text{ZX}$}
    \KwResult{optimized $P_l$, $P_r$, $\text{ZX}$}
    \For{c \KwIn $0 \dots q$}{
        \For{t \KwIn $0 \dots q$}{
            \If{c = t} {continue} 
            \If{$e(\text{ZX}, c, t) + e(P_l, c, t) + e(P_r, c, t) < 0 $}
            {   
            $P_l \gets \text{propagate}(P_l, c, t)$\;
            $P_r \gets \text{propagate}(P_r, c, t)$\;
            $ZX  \gets \text{propagate}(ZX, c, t)$\;
            }
        }   
    }
\end{algorithm}
Note that the computation of the Gaussian elimination for each step may be inefficient since it is polynomially bounded in $\mathcal{O}(q^3)$. Hence the overall process is in $\mathcal{O}(q^2(q^3 + nq^2)) = \mathcal{O}(q^5 + nq^4)$ if we consider every viable combination of $c$ and $t$.
We can improve this by filtering the search space beforehand. 
We note that the most significant benefit of propagating a CNOT through the parity map ($e_{\min}(P, c, t)$) is bounded by the cost of applying one CNOT.
This is the shortest path length between the control and the target. 
Algorithms like the Floyd-Warshall algorithm provide a lookup table for the path length. Hence one obtains the following upper bound:
\begin{equation}\label{eq:parity_upperbound}
        e_{\min}(P, c, t) \leq d(c, t)
\end{equation}
$d(c, t)$ is the shortest path length between the control and target.
One can see that this upper bound is valid by visualizing the process of Gaussian elimination as a sequence of CNOT gates on a quantum circuit. By adding a CNOT gate to the circuit, we can cancel out some CNOTs in the sequence. If that is not the case, we have found another step in the Gaussian elimination.
A consequence of Eq.~\eqref{eq:parity_upperbound} is that we can find a heuristic upper bound on the effect that we require on the ZX region:
\begin{equation*}
\begin{split}
e(c, t) & = e_{\min}(P_l, c, t) + e_{\min}(P_r, c, t) + e(\text{ZX}, c, t)\\
        & \leq 2 d(c, t) + e(\text{ZX}, c, t) \\
        & < 0
\end{split}
\end{equation*}
From the previous calculation, we can conclude that:
\begin{equation}
    e(ZX, c, t) < -2d(c, t)
\end{equation}
Lastly, we compute a Gaussian elimination once to optimize the parity region and propagate the CNOTs through the circuit if they show a negative effect. See algorithm~\ref{alg:fast_optimization} for outlining the optimization.
\begin{algorithm}
    \caption{Optimization using the derived heuristic.}\label{alg:fast_optimization}
    \SetKw{KwIn}{in}
    \KwData{current $P_l$, $P_r$, $\text{ZX}$}
    \KwResult{optimized $P_l$, $P_r$, $\text{ZX}$}
    \For{(c, t) \KwIn $\text{Steiner-Gauss}(P_l)$}{
        \If{$e(\text{ZX}, c, t) < -2d(c, t) $}
        {   
            $P_l    \gets \text{propagate}(P_l, c, t)$\;
            $P_r    \gets \text{propagate}(P_r, c, t)$\;
            $ZX     \gets \text{propagate}(ZX, c, t)$\;
        }
    }
    \For{(c, t) \KwIn $\text{Steiner-Gauss}(P_r)$}{
        \If{$e(\text{ZX}, c, t) < -2d(c, t) $}
        {   
            $P_l    \gets \text{propagate}(P_l, c, t)$\;
            $P_r    \gets \text{propagate}(P_r, c, t)$\;
            $ZX     \gets \text{propagate}(ZX, c, t)$\;
        }
    }
    \For{c \KwIn $0..q$}{
        \For{t \KwIn $0..q$}{
            \If{$e(\text{ZX}, c, t) < -2d(c, t) $}
            {   
                $P_l    \gets \text{propagate}(P_l, c, t)$\;
                $P_r    \gets \text{propagate}(P_r, c, t)$\;
                $ZX     \gets \text{propagate}(ZX, c, t)$\;
            }
        }   
    }
\end{algorithm}
We can see that per iteration of control or target, we require a computation of the effect on the ZX polynomial of $\mathcal{O}(nq^2)$ and per execution, we compute the recursive-Steiner-Gauss algorithm $\mathcal{O}(q^3)$. Hence the overall runtime is bounded by: $\mathcal{O}(nq^4 + q^3)$.
As the described process is heuristic, there may be occurrences where the parity region cost outweighs the benefit in the ZX region. Nevertheless, the heuristic performed well in practice, as observed in our evaluations.

\subsubsection{Regrouping of the ZX polynomial}
A key element of the proposed algorithm is splitting the ZX polynomial and optimizing the subregions. This procedure may result in suboptimal results if the ZX polynomial is not correctly adjusted before splitting. For example, consider the following polynomial:
\begin{equation*}
    \begin{ZX}
        \zxNone{}                           & \zxNone{}                           & \zxZ[]{\alpha_1}                    & \zxNone{}                           & \zxX[]{\beta_2}                     & \zxNone{}                           & \zxX[]{\beta_1}                     & \zxNone{}                           & \zxX[]{\beta_4}                     & \zxNone{}                           & \\
        \zxNone{}                           & \zxNone{}                           & \zxX[]{} \ar[u, bend right]         & \zxNone{}                           & \zxZ[]{} \ar[u, bend right]         & \zxNone{}                           & \zxZ[]{} \ar[u, bend right]         & \zxNone{}                           & \zxZ[]{} \ar[u, bend right]         & \zxNone{}                           & \\
        \zxNone{} \rar                      & \zxZ[]{} \ar[ru, bend right] \rar   & \zxNone{} \rar                      & \zxNone{} \rar                      & \zxNone{} \rar                      & \zxX[]{} \ar[ru, bend right] \rar   & \zxNone{} \rar                      & \zxNone{} \rar                      & \zxNone{} \rar                      & \zxNone{}                           & \\
        \zxNone{} \rar                      & \zxNone{} \rar                      & \zxNone{} \rar                      & \zxX[]{} \ar[ruu, bend right] \rar  & \zxNone{} \rar                      & \zxNone{} \rar                      & \zxNone{} \rar                      & \zxX[]{} \ar[ruu, bend right] \rar  & \zxNone{} \rar                      & \zxNone{}                           & \\
        \zxNone{} \rar                      & \zxZ[]{} \ar[ruuu, bend right] \rar & \zxNone{} \rar                      & \zxX[]{} \ar[ruuu, bend right] \rar & \zxNone{} \rar                      & \zxX[]{} \ar[ruuu, bend right] \rar & \zxNone{} \rar                      & \zxX[]{} \ar[ruuu, bend right] \rar & \zxNone{} \rar                      & \zxNone{}                           & \\
    \end{ZX}
\end{equation*}
One can see that both global optimizations will not show a positive effect, nor can one achieve a positive effect when splitting it in half and optimizing it further afterward. Nevertheless, when regrouping it such that the legs of the phase gadgets match, we can see that local optimization can achieve further effect by pulling out the two CNOTs.\begin{equation*}
    \begin{ZX}
        \zxNone{}                           & \zxNone{}                           & \zxZ[]{\alpha_1}                    & \zxNone{}                           & \zxX[]{\beta_1}                     & \zxNone{}                           & \zxX[]{\beta_2}                     & \zxNone{}                           & \zxX[]{\beta_4}                     & \zxNone{}                           & \\
        \zxNone{}                           & \zxNone{}                           & \zxX[]{} \ar[u, bend right]         & \zxNone{}                           & \zxZ[]{} \ar[u, bend right]         & \zxNone{}                           & \zxZ[]{} \ar[u, bend right]         & \zxNone{}                           & \zxZ[]{} \ar[u, bend right]         & \zxNone{}                           & \\
        \zxNone{} \rar                      & \zxZ[]{} \ar[ru, bend right] \rar   & \zxNone{} \rar                      & \zxX[]{} \ar[ru, bend right] \rar   & \zxNone{} \rar                      & \zxNone{} \rar                      & \zxNone{} \rar                      & \zxNone{} \rar                      & \zxNone{} \rar                      & \zxNone{}                           & \\
        \zxNone{} \rar                      & \zxNone{} \rar                      & \zxNone{} \rar                      & \zxNone{} \rar                      & \zxNone{} \rar                      & \zxX[]{} \ar[ruu, bend right] \rar  & \zxNone{} \rar                      & \zxX[]{} \ar[ruu, bend right] \rar  & \zxNone{} \rar                      & \zxNone{}                           & \\
        \zxNone{} \rar                      & \zxZ[]{} \ar[ruuu, bend right] \rar & \zxNone{} \rar                      & \zxX[]{} \ar[ruuu, bend right] \rar & \zxNone{} \rar                      & \zxX[]{} \ar[ruuu, bend right] \rar & \zxNone{} \rar                      & \zxX[]{} \ar[ruuu, bend right] \rar & \zxNone{} \rar                      & \zxNone{}                           & \\
    \end{ZX}
\end{equation*}
We conclude from this example that regrouping the ZX polynomial in subregions after optimization may be beneficial. To do this, we check for matching legs (legs that are on the same wire, regardless of their type) between neighboring phase gadgets in the following sense:
\begin{itemize}
    \item If two legs are on the same wire, we add $1$ to the score
    \item If no legs are on a wire, we add $-1$ to the score
    \item If there is a mismatch (one phase gadget has a leg present, one not), we add $-1$ to the score
\end{itemize}
With this score, we can define an order for phase gadgets a, b, and c in the following way:
\begin{itemize}
    \item Compute the score of a and c: $s_{ac}$
    \item Compute the score of a and b: $s_{ab}$
    \item The order is defined by $b <_{a} c = s_{ab} < s_{ac}$
\end{itemize}
In our example, we can see that $s_{\alpha_1\beta_1} = -2$ and $s_{\alpha_1\beta_2} = 1$. Hence, we can conclude that $\beta_1 <_{\alpha_1} \beta_2$. In other words, $\beta_2$ is more desirable to be a neighbor to $\alpha_1$ than $\beta_1$. We can incorporate this comparison into an insertion-sort-like structure as outlined in algorithm~\ref{alg:regrouping_sort}.
This process requires a computational overhead of $\mathcal{O}(n^2q)$.
\begin{algorithm}
    \caption{Regrouping of the ZX polynomial}\label{alg:regrouping_sort}
    \SetKw{KwIn}{in}
    \SetKw{KwAnd}{and}
    \KwData{ZX polynomial}
    \KwResult{ZX polynomial, that is grouped according to the score function}
    $\text{col} \gets 1$\;
    \While{$\text{col} < n-1$}{
        $\text{col}_{p} \gets \text{col} -1$\;
        $\text{col}_{c} \gets \text{col}$\;
        $\text{col}_{n} \gets \text{col} +1$\;
        \While{$\text{col}_{p} < n$ \KwAnd $ZX_{\text{col}_{p}} <_{ZX_{\text{col}_p}} ZX_{\text{col}_{n}}$}{
            swap($\text{col}_c$, $\text{col}_{n}$)\;
            $\text{col}_{p} \gets \text{col}$\;
            $\text{col} \gets \text{col}_n$\;
            $\text{col}_{n} \gets \text{col}+1$\;
        }
        $\text{col} \gets \text{col} + 1$
    }
\end{algorithm}

\subsubsection{Splitting}
\begin{figure*}[ht]
    \centering
    \begin{quantikz}[row sep=0.5em]
        \lstick{}& \qw      & \gate[5, nwires=3]{I}             & \qw       & \gate[5, nwires=3]{\text{ZX (left)}} & \qw    & \gate[5, nwires=3]{\text{I}} & \qw    & \gate[5, nwires=3]{\text{ZX (right)}} & \qw    & \gate[5, nwires=3]{P_r} & \qw    & \qw &\\
        \lstick{}& \qw      &                                   & \qw       &                                      & \qw    &                              & \qw    &                                       & \qw    &                         & \qw    & \qw &\\
        \lstick{}& \vdots   &                                   & \vdots    &                                      & \vdots &                              & \vdots &                                       & \vdots &                         & \vdots &     &\\
        \lstick{}& \qw      &                                   & \qw       &                                      & \qw    &                              & \qw    &                                       & \qw    &                         & \qw    & \qw &\\
        \lstick{}& \qw      &                                   & \qw       &                                      & \qw    &                              & \qw    &                                       & \qw    &                         & \qw    & \qw &\\
    \end{quantikz}
    \caption{Sample generation of a matrix representing the binary legs of a phase gadget.}\label{fig:zx_region_split_split}
\end{figure*}
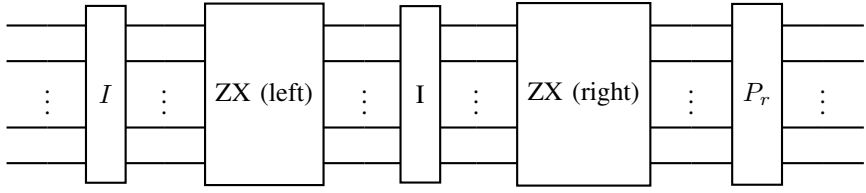

Since we want to execute our optimization strategy on local regions of the ZX polynomial, we developed a divide-and-conquer approach to optimize sub-regions after they had been optimized globally. We organize the ZX polynomial by ``padding'' it with two parity maps, as outlined in Figure~\ref{fig:zx_region_split}.
Next, we aim to pull out as many CNOTs as possible, sort the ZX polynomial, and split it in half, padding the gap with an identity parity map (see Figure \ref{fig:zx_region_split_split}). 
We then execute \emph{divide\_fast} again until we reach a size of the ZX polynomial $\leq 2$. In that case, we optimize this region and return it. This results in an alternating set of ZX and parity regions, which we can individually convert to a quantum circuit via the synthesis method detailed in \cite{PauliOpt} or the \textit{recursive-Steiner-Gauss} algorithm, respectively. See algorithm~\ref{alg:splitting_algorithm} for an outline of the general algorithm.
\begin{algorithm}
    \caption{Overall process of the divide and conquer algorithm}\label{alg:splitting_algorithm}
    \SetKw{KwIn}{in}
    \SetKw{KwAnd}{and}
    \SetKw{Require}{Require:}
    \KwData{ZX polynomial, Architecture}
    \KwResult{qc: A Quantum Circuit that represents the ZX polynomial routed on the Architecture}
    
    \SetKwFunction{FDivide}{synthesize}
      \SetKwProg{Fn}{Function}{:}{}

    \Require $\text{ZX}$ the ZX polynomial\;
    \Require $P_l$ left parity of the ZX polynomial\;
    \Require $P_r$ right parity of the ZX polynomial\;
    
    \Fn{\FDivide{$P_l$, $\text{ZX}$, $P_r$}}{
        $\text{ZX} \gets \text{regroup}(\text{ZX})$\;
        $P_l, \text{ZX}, P_r \gets \text{optimize}(P_l, \text{ZX}, P_r)$\;
        $\text{ZX}_l, \text{ZX}_r \gets \text{split}(\text{ZX})$
        $P_c \gets I$\;
        $\text{region}_l \gets \text{synthesize}(P_c, \text{ZX}_r, P_r)$\;
        $\text{region}_r \gets \text{synthesize}(P_l, \text{ZX}_l, P_c)$\;
        \KwRet{$\text{region}_l + \text{region}_r$}\;
    }
    $\text{ZX} \gets \text{regroup}(\text{ZX})$\;
    $P_{ol}, \text{ZX}, P_{or} \gets \text{optimize}(P_l, \text{ZX}, P_r)$\;
    $P_l \gets I$\;
    $P_r \gets I$\;
    $\text{region} \gets \text{synthesize}(P_c, \text{ZX}, P_r)$\;
    $\text{region} \gets [P_{ol}] + \text{region} + [P_{or}]$\;
    $\text{qc} \gets \text{regionToCircuit}(\text{region})$\;

\end{algorithm}

\subsubsection{Computational cost}\label{sec:synthesis_computational_cost}
We provide two variants for optimization, one where we compute the \emph{recursive Steiner-Gauss} algorithm at each step to evaluate the costs (\emph{divide\_gauss} in the following) and one where we assumed the cost by an upper bound (\emph{divide\_fast} in the following). Hence we can then find a runtime of $\mathcal{O}(q^5 + nq^4)$ for \emph{divide\_fast} and a runtime of $\mathcal{O}(q^3 + nq^4) = \mathcal{O}(nq^4)$ for \emph{divide\_gauss}. Additionally, we sort the ZX Polynomial at each recursion step, which yields a runtime of $\mathcal{O}(qn^2)$.
Our algorithm consists of a split of the ZX Polynomial at one recursion step plus an additional optimization process that differs from \emph{divide\_fast} to \emph{divide\_gauss} in terms of runtime. Hence, we can describe the complexity behavior of these two operations by the following recurrence relations:

\begin{equation*}
T(n) = \begin{cases}
    g(n)                                                           & n \leq 2 \\
    T(\lfloor n/2 \rfloor ) + T(\lceil n / 2 \rceil) + g(n)    & \text{else}
\end{cases}
\end{equation*}
We can then utilize the Akara-Bazi theorem~\cite{Akra1998}, using the worst-case complexity of the base cases as $g(n) = nq^4 + q^3 + qn^2$ for \emph{divide\_fast}, and $g(n) = q^5 + nq^4 + n^2q$ for \emph{divide\_gauss}.
We can easily determine that $p=1$ for both algorithms and directly obtain the asymptotic behavior for the two cases. $q$ is constant for this analysis as the number of qubits remains constant for a given circuit. For \emph{divide\_fast}:
\begin{align*}
    T(n) &\in \Theta \left( n^1 (1+ \int_{1}^{n} \frac{uq^4 + q^3 + qu^2}{u^{p+1}}) \,du\ \right) \\
         & = \Theta\left(n (1 + q^4\log (n) + q^3 + qn - \frac{q^3}{n} - q)\right) \\
         & = \Theta\left(n + q^4 n \log (n) + q^3n + qn^2 - q^3 - qn\right) \\
         & = \mathcal{O}\left( q^4n \log (n) + qn^2\right)
\end{align*}
Analogously for \emph{divide\_gauss}:
\begin{align*}
    T(n)    &\in \Theta \left( n^1 (1+ \int_{1}^{n} \frac{q^5 + uq^4 + u^2q}{u^{p+1}}) \,du\ \right) \\
            & = \Theta \left( n + q^4n \log (n) + qn^2 + q^5n - qn - q^5 \right) \\
            & = \mathcal{O} (q^4 n \log (n) + qn^2 + nq^5)
\end{align*}
We can then conclude the following runtimes for the introduced algorithms:
\begin{itemize}
    \item \emph{divide\_fast}: $\mathcal{O}\left( q^4n \log (n) + qn^2\right)$
    \item \emph{divide\_gauss}: $\mathcal{O} \left(q^4 n \log (n) + qn^2 + nq^5\right)$
\end{itemize}
While the change in computational complexity does not seem significant as $n \gg q$, we do note that additionally caching CNOT placements for ZX phase gadgets in \textit{divide\_fast} significantly reduced runtime in practice.
\section{Numerical Experiments}
We compare our approach to PauliOpt, pyzx, and the algorithm of Cowtan et al.~\cite{Cowtan2020} (tket in the following). 
First, pyzx and tket were routed by the default procedure of the tket framework~\cite{cowtan_et_al:LIPIcs:2019:10397}. 
Next, we used the \emph{cnot\_count} property for PauliOpt to get the routed CNOT count. 
Finally, we counted the CNOTs in the resulting quantum circuit for the other algorithms. 
For PauliOpt, we used 4800 iterations and the following annealing schedule: (linear, 1.5, 0.1). 
The other algorithms were run with default parametrization.
As a measure we use the CNOT reduction count: $\frac{\#cx_{o} - \#cx_i}{\#cx_i}$, here $\#cx_{o}$ is the number of CNOTs of the resulting circuit, $\#cx_i$ the number of the naive synthesized circuit.
Four experimental setups were designed and executed. 
The first compared the execution time and performance of \emph{divide\_fast} and \emph{divide\_gauss} to confirm that acceptable performance was retained when using the introduced heuristic. 
Then, we reproduced results from PauliOpt and compared them to our approach. 
Next, we compared the performance of PauliOpt, pyzx, tket, and \emph{divide\_fast}, first using a random ansatz and then the well-known MaxCut QAOA ansatz.

\subsection{Comparison of \emph{divide\_fast} and \emph{divide\_gauss}}
In section~\ref{sec:synthesis_computational_cost}, we described two different methods for optimizing the ZX regions after a split. Their main difference results from the fact that \emph{divide\_gauss} uses a Gaussian elimination at each iteration of the optimization process, while \emph{divide\_fast} is based on a heuristic upper bound. We conducted experiments to compare their ability to synthesize a ZX polynomial and their practical runtime, which we measured in seconds. For this experiment, we varied the number of qubits (\emph{n\_qubits}) in $\{4, 5, 6\}$. Furthermore, we used random ZX polynomials with phase gadgets in $\{10, 30, 50, 70, 90\}$ and set the maximum allowed legs to four. Finally, we used a fully connected, a line, and a circle architecture. We ran each configuration of the parameters above 20 times.
\begin{table}
    \centering
    \begin{tabular}{ll|ll}
        \toprule
        n\_qubits   &  algorithm  &  time\_diff $[s]$       &  two\_qubit\_reduction $[\%]$ \\
        \midrule
        4       & divide\_fast  &       $0.19\pm 0.10$  &                $60.49\pm10.97$ \\
                & divide\_gauss &       $0.79\pm 0.37$  &                $63.06\pm10.35$ \\
        5       & divide\_fast  &       $0.32\pm 0.19$  &                $45.28\pm11.66$ \\
                & divide\_gauss &       $2.51\pm 1.39$  &                $49.02\pm10.05$ \\
        6       & divide\_fast  &       $0.46\pm 0.27$  &                $40.74\pm09.60$ \\
                & divide\_gauss &       $6.30\pm 3.53$  &                $44.92\pm08.72$ \\
        \bottomrule
    \end{tabular}
    \caption{Time difference and two qubit reduction of \emph{divide\_fast} and \emph{divide\_gauss}}\label{tab:fast_gauss_evaluations}
\end{table}
Our results can be found in Table~\ref{tab:fast_gauss_evaluations}. As stated in section~\ref{sec:synthesis_computational_cost}, it shows that while the complexity classes are equivalent, the two algorithms differ significantly regarding runtime. Furthermore, in terms of accuracy, we see a constant drop from \emph{divide\_fast} of around four percent, which we determine to be an acceptable loss in performance for the significant speedup achieved. Hence, for the remaining comparisons in this work, we provide the performance of \emph{divide\_fast}.

\subsection{Reproducing the results by PauliOpt}
\begin{figure}[ht!]
    \centering
    \includegraphics*[width=0.8\linewidth]{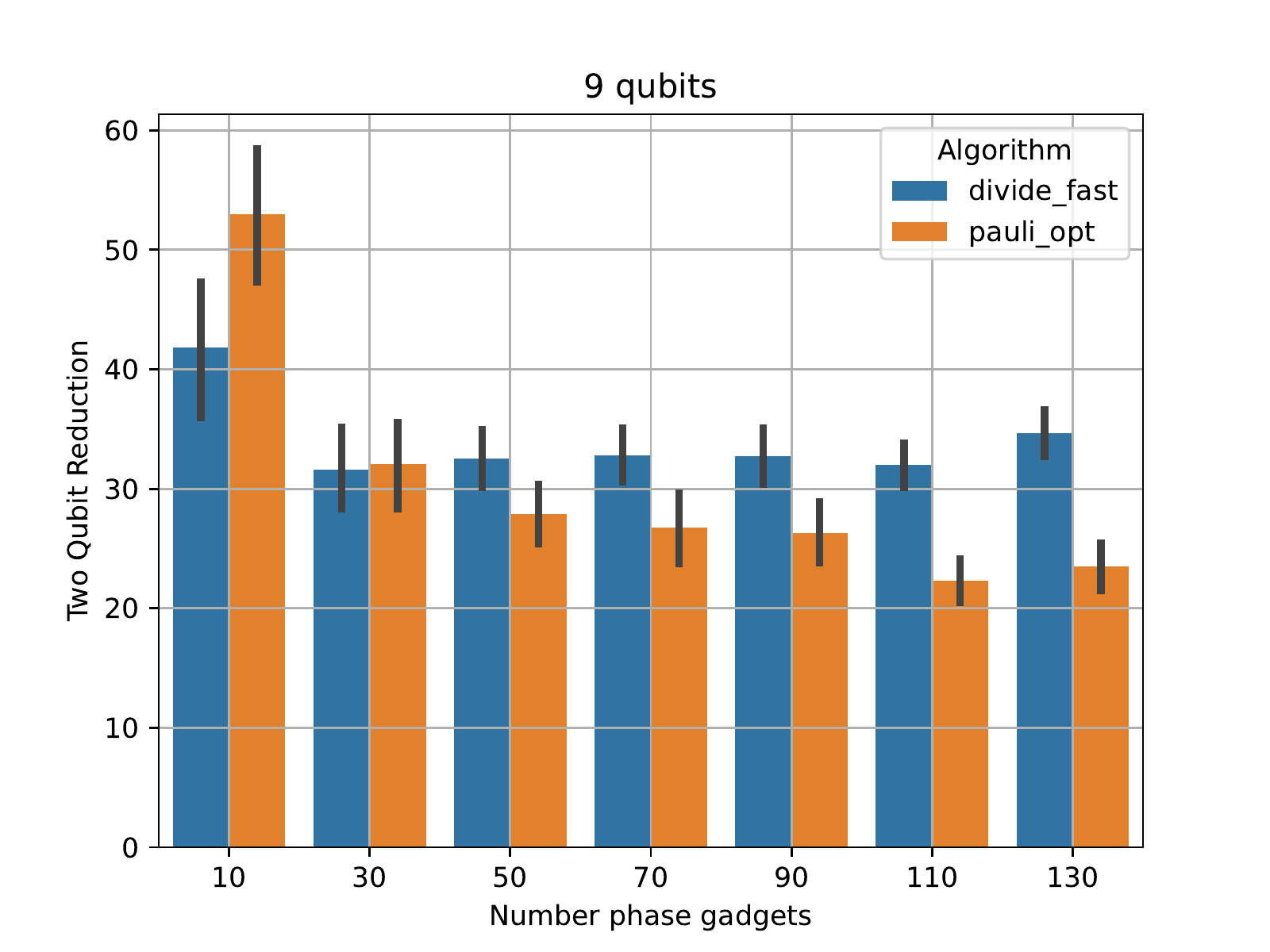}
    \includegraphics*[width=0.8\linewidth]{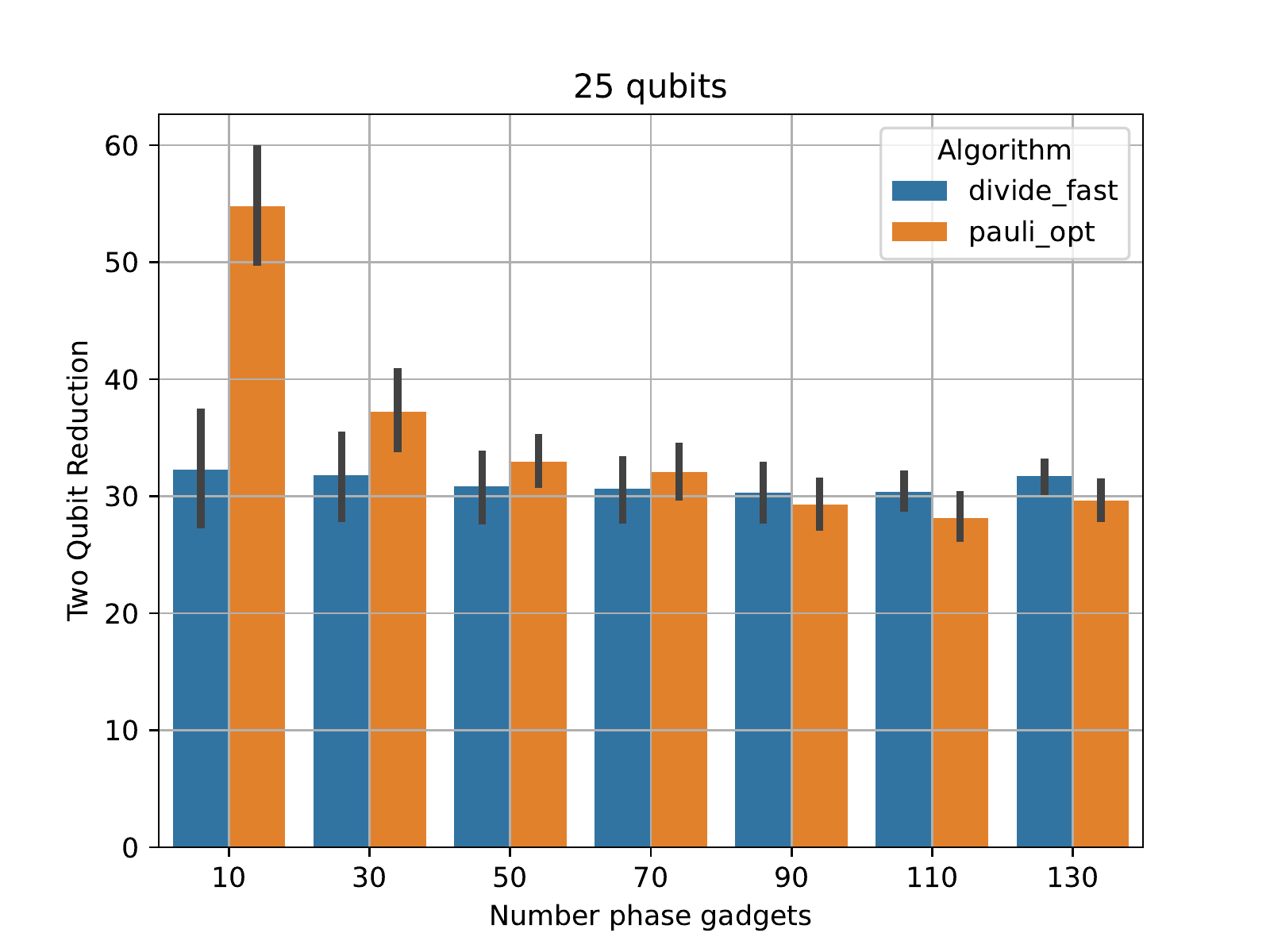}
    \caption{Reproduction of results from PauliOpt compared to \textit{divide\_fast} with respect the number of phase gadgets in a ZX polynomial for 9 and 25 qubits.
    }\label{fig:reproduce_pgs}
\end{figure}

For reproducing the results of Gogioso et al.~\cite{PauliOpt}, we kept the maximum legs per phase gadgets towards four and used a square architecture. We randomly generated phase polynomials with $\{10, 30, 50, 70, 90, 110, 130\}$ phase gadgets over a range of qubits $\{ 9, 16, 25 \}$.
We generated 20 ZX polynomials for each combination and optimized each randomly generated circuit using PauliOpt and \emph{divide\_fast}. 
The results for 9 and 25 qubits are presented in Figure~\ref{fig:reproduce_pgs}. One can observe the numerical values of the experiment in Table~\ref{tab:pgs_qubits_gauss_fast}.
Firstly, we see that, as expected, both algorithms performance degenerates with increasing phase gadgets.
Next, while \emph{divide\_fast} is outperformed at a lower number of phase gadgets, it does demonstrate better scaling capabilities with increased phase gadgets. 
This supports the hypothesis that partitioning the ZX polynomial during optimization significantly affects larger polynomials. 
The results hint that incorporating the partitioning method introduced in this work may improve the performance of PauliOpt. 
\begin{table}
    \centering
    \begin{tabular}{llll}
        \toprule
        n\_qubits & n\_pgs & algorithm & two\_qubit\_reduction $[\%]$ \\
        \midrule
        9  & 10  & divide\_fast &      41.82$\pm$14.27 \\
           &     & pauli\_opt &      52.96$\pm$13.63 \\
           & 30  & divide\_fast &        31.6$\pm$8.19 \\
           &     & pauli\_opt &        32.1$\pm$8.58 \\
           & 50  & divide\_fast &       32.51$\pm$5.84 \\
           &     & pauli\_opt &       27.89$\pm$5.92 \\
           & 70  & divide\_fast &       32.77$\pm$5.27 \\
           &     & pauli\_opt &       26.74$\pm$7.35 \\
           & 90  & divide\_fast &       32.72$\pm$5.81 \\
           &     & pauli\_opt &       26.31$\pm$6.11 \\
           & 110 & divide\_fast &       31.98$\pm$4.26 \\
           &     & pauli\_opt &       22.34$\pm$4.11 \\
           & 130 & divide\_fast &        34.64$\pm$4.4 \\
           &     & pauli\_opt &       23.52$\pm$4.77 \\
        16 & 10  & divide\_fast &      34.46$\pm$15.48 \\
           &     & pauli\_opt &      51.04$\pm$12.84 \\
           & 30  & divide\_fast &       34.47$\pm$9.66 \\
           &     & pauli\_opt &      39.11$\pm$10.21 \\
           & 50  & divide\_fast &       31.18$\pm$5.89 \\
           &     & pauli\_opt &       29.55$\pm$5.69 \\
           & 70  & divide\_fast &       31.29$\pm$5.99 \\
           &     & pauli\_opt &       28.54$\pm$5.99 \\
           & 90  & divide\_fast &       29.23$\pm$6.47 \\
           &     & pauli\_opt &       26.21$\pm$6.49 \\
           & 110 & divide\_fast &       29.95$\pm$3.42 \\
           &     & pauli\_opt &       25.57$\pm$4.51 \\
           & 130 & divide\_fast &        30.93$\pm$5.1 \\
           &     & pauli\_opt &       26.26$\pm$6.33 \\
        25 & 10  & divide\_fast &       32.3$\pm$11.42 \\
           &     & pauli\_opt &      54.78$\pm$11.93 \\
           & 30  & divide\_fast &       31.82$\pm$7.99 \\
           &     & pauli\_opt &        37.2$\pm$7.39 \\
           & 50  & divide\_fast &       30.82$\pm$6.73 \\
           &     & pauli\_opt &       32.96$\pm$4.61 \\
           & 70  & divide\_fast &       30.62$\pm$6.32 \\
           &     & pauli\_opt &        32.06$\pm$4.8 \\
           & 90  & divide\_fast &        30.27$\pm$5.4 \\
           &     & pauli\_opt &        29.3$\pm$4.54 \\
           & 110 & divide\_fast &        30.4$\pm$3.34 \\
           &     & pauli\_opt &       28.14$\pm$4.28 \\
           & 130 & divide\_fast &       31.73$\pm$2.88 \\
           &     & pauli\_opt &       29.64$\pm$3.75 \\
        \bottomrule
        \end{tabular}
    \caption{Numerical results for the evaluation of 9, 16 and 25 qubits for a selection of 10, 30, 50, 70, 90, 110, and 130 number of phase gadgets.}\label{tab:pgs_qubits_gauss_fast}
\end{table}

\subsection{Evaluation for random circuits}
\begin{figure}[ht!]
    \includegraphics*[width=0.8\linewidth]{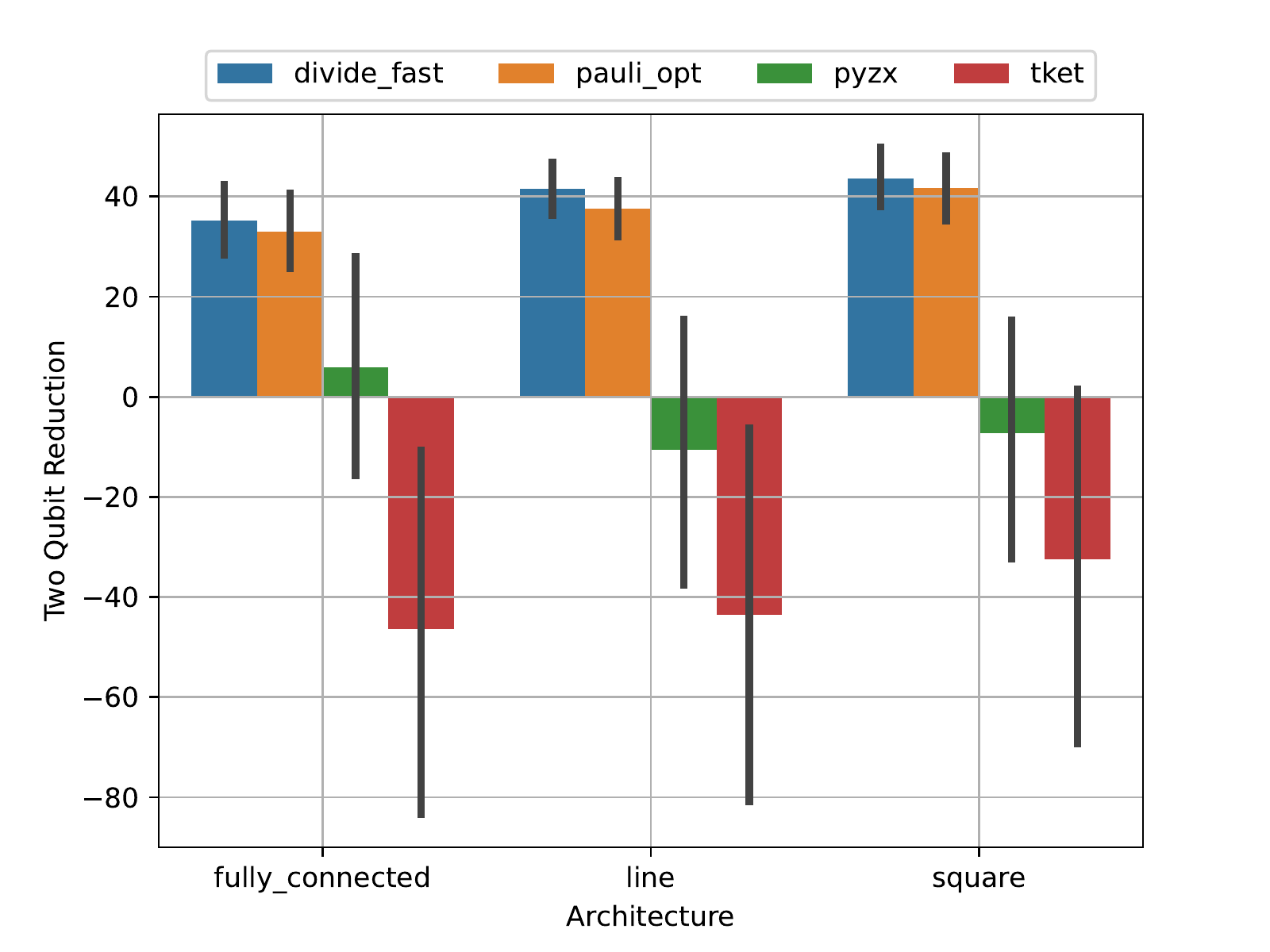}
    \caption{Evaluation of random ZX polynomials for PauliOpt, pyzx, tket and \emph{divide\_fast}}\label{fig:random_evaluation}
\end{figure}

We generated a more extensive set of random circuits with fewer qubits to see the behavior of \emph{divide\_fast} compared to other optimization algorithms. 
We varied the number of qubits (\emph{n\_qubits}), in the range $\{4, 9, 16\}$, the number of phase gadgets (\emph{n\_pgs}) in the range $\{15, 32, 64, 128\}$. 
Next, we changed the parameter to specify the maximum number of legs (\emph{max\_legs}) as either $3$ or $4$. 
Lastly, we tried synthesizing the ZX polynomial towards a line, square, and fully connected architecture. 
For random circuits, we found that \emph{divide\_fast} scaled the best for varying architectures (see Figure~\ref{fig:random_evaluation}).
We can see that pyzx and tket perform worse in the random scenario. This may reveal a particular weakness of these algorithms to randomly generated circuits, which do not significantly impact the performance of PauliOpt or \emph{divide\_fast}.
While this may be due to SWAPs added during routing, we also see a difference in performance in the case of fully connected architecture, where no SWAPs were added.
Another exciting aspect of the proposed algorithm is the stability among architectures. While we can see that switching from a line towards a fully-connected architecture may impact the performance of pyzx, the performance of \emph{divide\_fast} remains consistent.
\begin{figure}[ht!]
    \includegraphics*[width=0.9\linewidth]{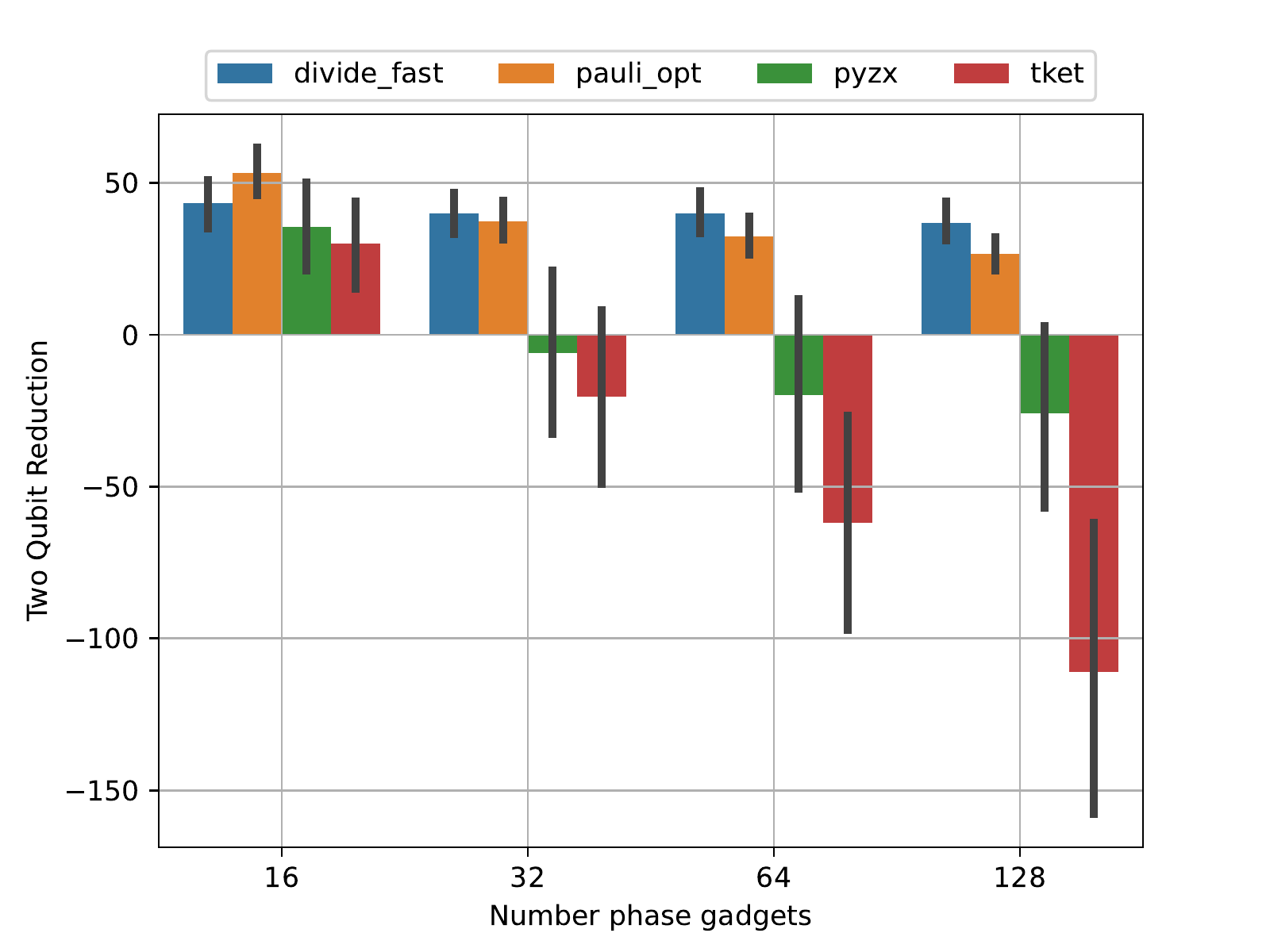}
    \includegraphics*[width=0.9\linewidth]{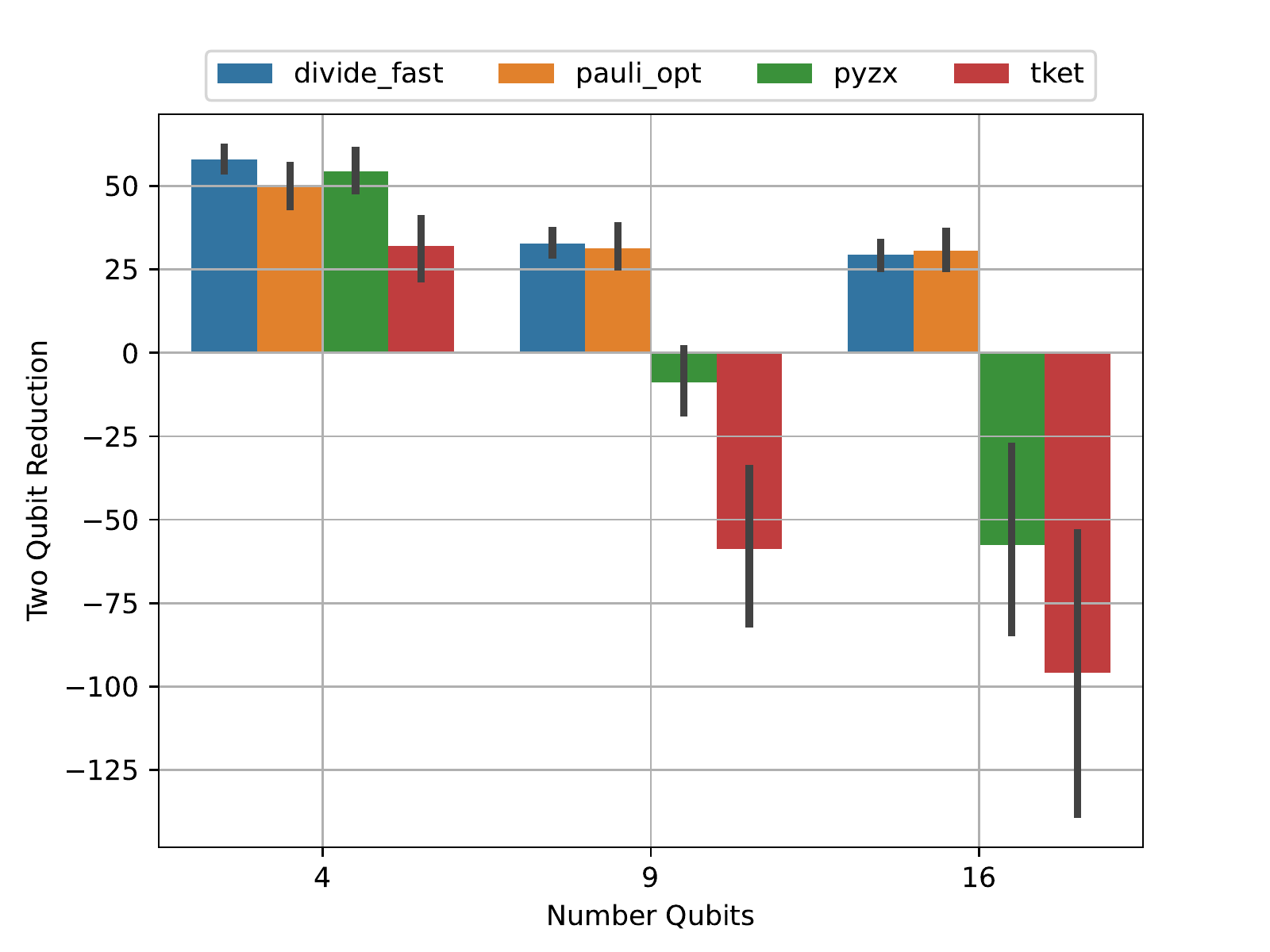}
    \caption{Evaluation of random ZX polynomials for PauliOpt, pyzx, tket and \emph{divide\_fast}}\label{fig:random_evaluation_qubits_pgs}
\end{figure}
When varying the number of phase gadgets and qubits, we found that overall the performance of tket and pyzx degenerated for large random ZX polynomials. At the same time, both PauliOpt and our approach tended to show consistent performance, as seen in Figure~\ref{fig:random_evaluation_qubits_pgs}.

\subsection{Evaluation for MaxCut}
\begin{figure}
    \includegraphics*[width=0.8\linewidth]{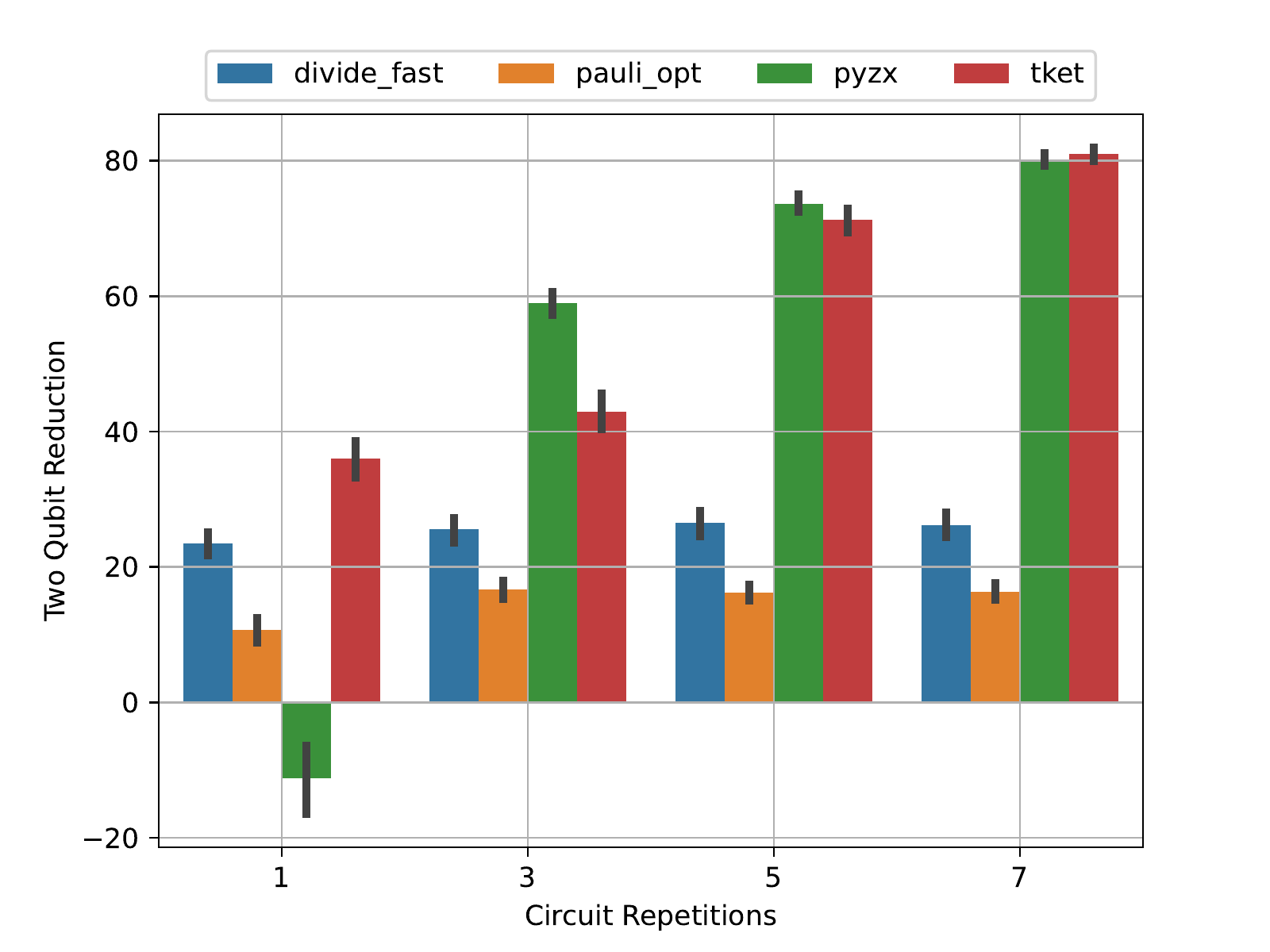}
    \includegraphics*[width=0.8\linewidth]{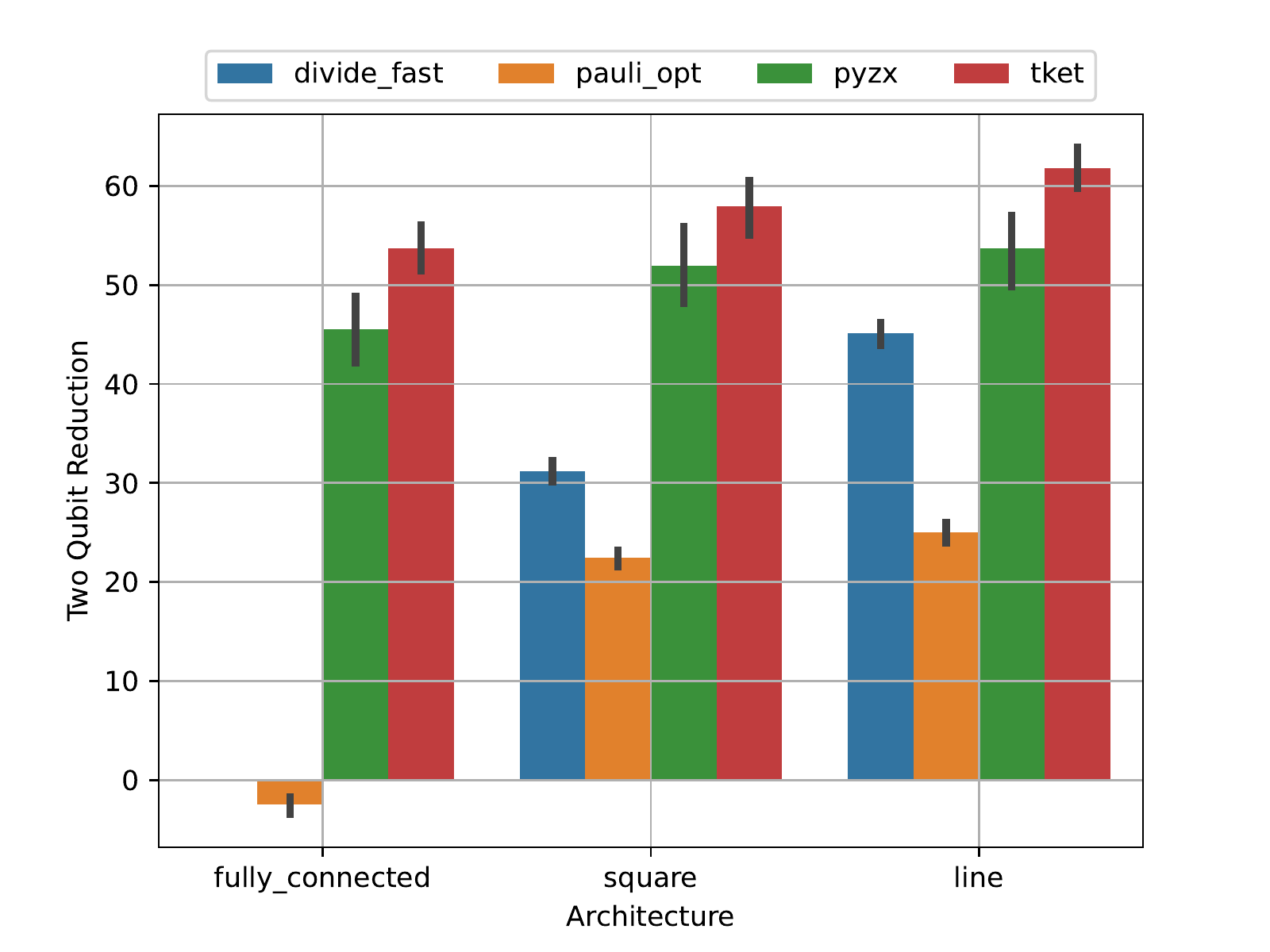}
    \caption{Evaluation results for MaxCut circuits for number of circuit repitions and architectures.}\label{fig:max_cut_evaluation}
\end{figure}

\begin{figure}
    \includegraphics*[width=0.8\linewidth]{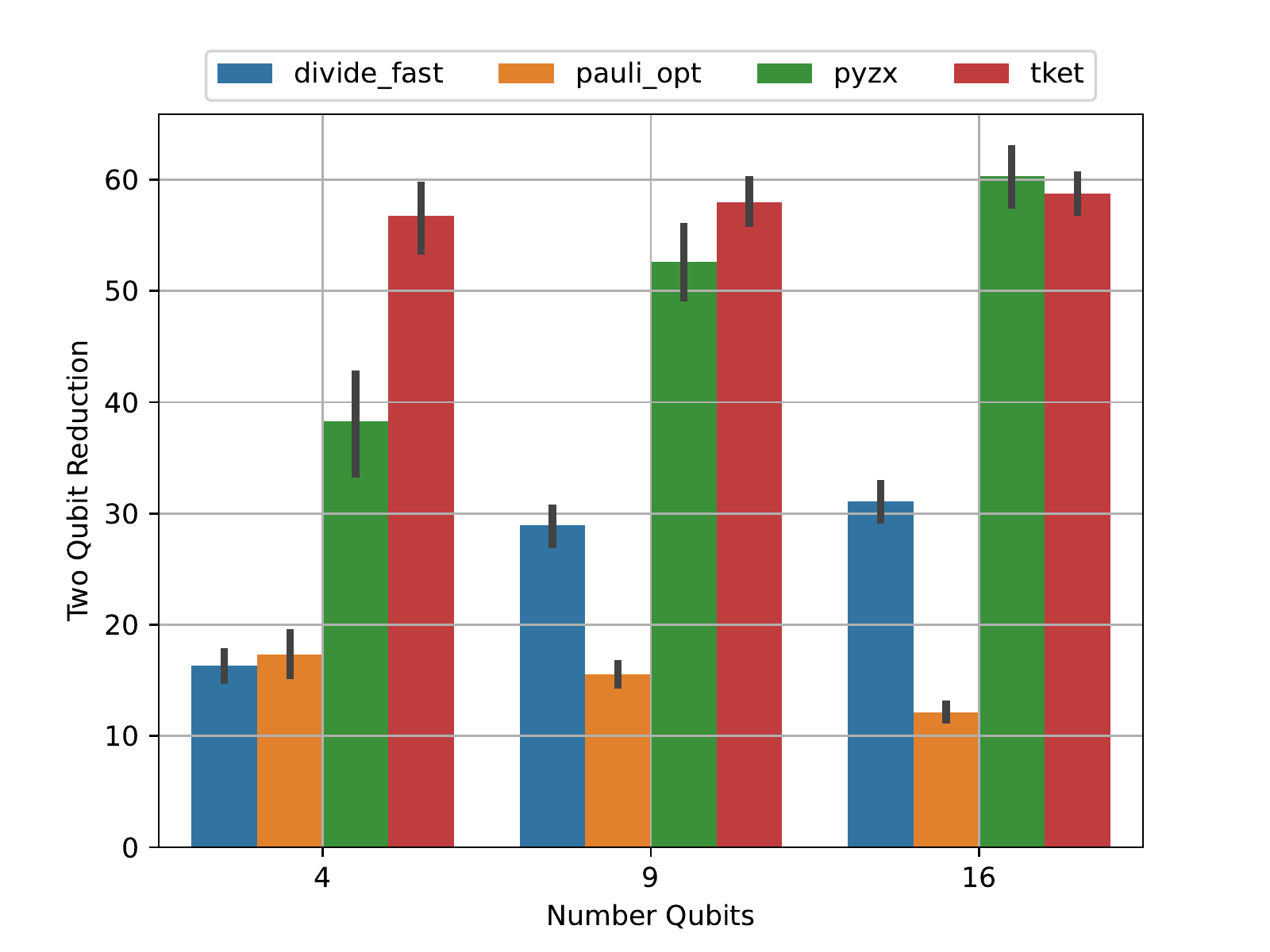}
    \caption{Evaluation results for MaxCut circuits with respect to the number of qubits}\label{fig:max_cut_evaluation_2}
\end{figure}

For the MaxCut ansatz, we used Erd\H{o}s-R\'{e}nyi graphs~\cite{Erdos1969} to generate the ZX polynomial to be optimized. 
We decided to remove the leading Hadamard gates, which do not directly affect the CNOT count of the circuit but provides more straightforward structure for optimization since the circuit is balanced in each repetition. 
In a real-world example, one could reintroduce the leading Hadamard gates, leaving the unitary unchanged.
We varied the number of vertices (\emph{n\_vertices}) \footnote{In the standard QAOA Ansatz, the number of vertices in the graph corresponds to the number of qubits in the quantum circuit} in the range $\{4, 9, 16\}$, for the edge probability (\emph{p\_edge}) we used the following parameters: $\{0.5, 0.7, 0.9\}$, the different layer repetitions where altered by the following parameters: $\{1, 3, 5, 7\}$. 
Lastly, we used a fully connected line and a square architecture as in the random evaluation.
Here one can observe (see Figure~\ref{fig:max_cut_evaluation}) that both approaches PauliOpt and our synthesis algorithms, outperformed both tket and pyzx. 
We observe in the optimized polynomials that tket and pyzx can reduce the overall circuit depth with either a smart way to expose multiple CNOTs towards a Clifford simplification algorithm or clever SWAP placements in a structured circuit.
This observation stems from the vastly improved performance of these two algorithms by adding circuit repetitions; the optimization algorithms can remove multiple CNOTs throughout the repeated circuits.
This is understandable as our introduced algorithm partitions the circuit; this reduces the problem size and may hinder global optimizations. \emph{divide\_fast} tends to perform comparably better for sparser architectures (compare the line and square architecture towards the fully connected one). This is another indicator that being hardware agnostic in the optimization process of a transpiling algorithm can be beneficial.
Interestingly our divide and conquer approach tends to scale better than PauliOpt with increased qubit count (see Figure~\ref{fig:max_cut_evaluation_2}).
This may indicate that a heuristic-based search may scale better than the non-deterministic approach used by PauliOpt.

\section{Conclusion}
In this work, we have extended the algorithm by Gioso et al.~\cite{PauliOpt} by introducing a heuristic-based optimization strategy with precise complexity bounds and a partitioning approach, demonstrating better scaling with increased qubit count and phase gate depth.
We have additionally evaluated these algorithms against state-of-the-art circuit optimization algorithms, examining the performance of each algorithm with randomly generated and structured circuits.
While the state-of-the-art optimization strategies outperform PauliOpt and \emph{divide\_fast} on structured circuits, these algorithms do not perform well with randomly generated circuits.
The results further highlight a reliance on the examined state-of-the-art optimization strategies for suitable mapping and routing post-processing steps.
This provides a strong argument for algorithms that can synthesize circuits in a architecture-aware manner as opposed to optimization algorithms that require a second mapping and routing step. 
The overall performance of PauliOpt and \emph{divide\_fast} is not strongly impacted when moving between architectures of varying connectivity.
As a follow-up on this work, we propose extending the synthesis algorithm to allow for Pauli polynomials, which may provide additional optimization possibilities. Additionally, we would like to provide our algorithms in the PauliOpt library since they already provide a general and intuitive way of writing ZX Polynomials.

%
%
%
\appendix




\section*{Proof of the $\pi$-commutation rule}\label{sec:proof_pi_commutation}
In this section, we will provide proof for Eq.~\eqref{eq:phase_commutation_pi_z} by showing it for an X Phase gadget with phase $\pi$ and a Z phase gadget with an arbitrary phase. Due to the color duality in the ZX-Calculus, the opposite case as in Eq.~\eqref{eq:phase_commutation_pi_x} holds too.
We used the following rules from~\cite{Wetering_ZXCalc}:
\begin{center}
\setlength{\tabcolsep}{0pt}
\renewcommand{\arraystretch}{0} 
\tikzexternaldisable
\begin{tabularx}{1.00\linewidth}{XX}
	\begin{equation}\label{eq:b}
		\begin{ZX}[math baseline=cent]
			\zxNone{} \rar & \zxZ{}\rar\ar[ddrr]   & \zxNone{} \rar & \zxX{}\rar  & \zxNone{} \\
			\zxNone[a=cent]{} \\
			\zxNone{} \rar & \zxZ{}\rar\ar[uurr]   & \zxNone{} \rar & \zxX{}\rar  & \zxNone{} \\
		\end{ZX} = 
		\begin{ZX}[math baseline=cent]
			\zxNone{} &             &                           & \zxNone{} \\
			\zxNone{} & \zxX{}\ar[ul, bend right]\ar[dl, bend left] \rar & \zxZ{}\ar[ru, bend left] \ar[rd, bend right]{}    &           \\
			\zxNone{} &             &                           & \zxNone{} \\
		\end{ZX}
	\end{equation} &
	\begin{equation}\label{eq:minus_pi}
		\begin{ZX}[math baseline=cent]
			\zxNone{} \rar & \zxX[a=cent]{\pi}\rar & \zxZ{\alpha} \rightManyDots{} \\
		\end{ZX} = 
		\begin{ZX}[math baseline=cent]
			\zxNone{} \rar & \zxNone{} \rar & \zxZ{-\alpha} \rightManyDots{} \\
		\end{ZX}
	\end{equation} \\[-1em]
	\begin{equation}\label{eq:pi_copy_split}
		\begin{ZX}[math baseline=cent]
			\zxNone{} \\
			\zxNone[a=cent]{} \rar & \zxX{\pi} \rar & \zxZ{} \rightManyDots{}  & \zxNone{} \\
			\zxNone{}
		\end{ZX} = 
		\begin{ZX}[math baseline=cent]
			\zxX{\pi} \rar & \zxNone{} \ar[dd,3 vdots] \\
			\zxNone[a=cent]{} \\
			\zxX{\pi} \rar & \zxNone{} \\
		\end{ZX}
	\end{equation} &
	\begin{equation}\label{eq:comp}
		\begin{ZX}[math baseline=cent]
			\zxNone[a=cent]{} \rar & \zxZ{} \ar[r,o'] \ar[r,o.] & \zxX{}\rar  & \zxNone{}
		\end{ZX} = 
		\begin{ZX}[math baseline=cent]
			\zxNone[a=cent]{} \rar & \zxZ{} & \zxX{}\rar  & \zxNone{}
		\end{ZX}
	\end{equation} \\[-0.7em]
	\multicolumn{2}{>{\hsize=\dimexpr2\hsize+2\tabcolsep+\arrayrulewidth\relax}X}{\begin{equation}\label{eq:fusion}
		\begin{ZX}[math baseline=cent]
			\leftManyDots{} \zxX[a=cent]{\alpha}\rar & \zxX{\beta} \rightManyDots{} \\
		\end{ZX} = 
		\begin{ZX}[math baseline=cent]
			\leftManyDots{} \zxX[a=cent]{\alpha + \beta} \rightManyDots{}
		\end{ZX}
	\end{equation}}
\end{tabularx}
\tikzexternalenable
\setlength{\tabcolsep}{6pt}
\end{center}
We start similarly as \cite{Yeung2020} to swap the shared legs of the two-phase gadgets:~\footnote{Note that the non-shared can be swapped directly}
\begin{equation*}
	\begin{aligned}
		&\begin{ZX}
			\zxNone{}      &                			& \zxZ[a=pgp1]{\alpha_i}&
			\zxNone{}      &                			& \zxX[a=pgp1]{\pi}\\
			\zxNone{}      & 					& \zxX[a=pg1]{}	\ar[u, bend right]&
			\zxNone{}      & 					& \zxZ[a=pg1]{}	\ar[u, bend right]\\
			\zxNone{} \\
			\zxNone{} \rar & \zxZ{} \ar[ruu, bend right] \rar 	& \zxNone{}\rar&
			\zxNone{} \rar & \zxX{} \ar[ruu, bend right] \rar 	& \zxNone{}\\
			\zxNone{} \rar & \zxZ{} \ar[ruuu, bend right] \rar 	& \zxNone{}\rar&
			\zxNone{} \rar & \zxX{} \ar[ruuu, bend right] \rar 	& \zxNone{}\\
			\zxNone{}	& \vdots{}				&	&
			\zxNone{}	& \vdots{}				&	\\
			\zxNone{} \rar & \zxZ{}\ar[ruuuuu, bend right] \rar	& \zxNone{}\rar&
			\zxNone{} \rar 	& \zxX{}\ar[ruuuuu, bend right] \rar	& \zxNone{}\\
		\end{ZX}
		\overset{\eqref{eq:b}}{=}
		\begin{ZX}
			\zxNone{}      &                			& \zxZ[a=pgp1]{\alpha_i}&
			\zxNone{}      &                			& \zxX[a=pgp1]{\pi}\\
			\zxNone{}      & 					& \zxX[a=pg1]{}	\ar[u, bend right] \ar[rrr] &
			\zxNone{}      & 					& \zxZ[a=pg1]{}	\ar[u, bend right]\\
			\zxNone{} \\
			\zxNone{} \rar & \zxX{} \ar[rrrruu] \rar 		& \zxNone{}\rar&
			\zxNone{} \rar & \zxZ{} \ar[lluu] \rar 			& \zxNone{}\\
			\zxNone{} \rar & \zxZ{} \ar[ruuu, bend right] \rar 	& \zxNone{}\rar&
			\zxNone{} \rar & \zxX{} \ar[ruuu, bend right] \rar 	& \zxNone{}\\		
			\zxNone{}	& \vdots{}				&	&
			\zxNone{}	& \vdots{}				&	\\
			\zxNone{} \rar & \zxZ{}\ar[ruuuuu, bend right] \rar	& \zxNone{}\rar&
			\zxNone{} \rar 	& \zxX{}\ar[ruuuuu, bend right] \rar	& \zxNone{}\\
		\end{ZX}
		\overset{\eqref{eq:b}}{=}\\
		&\begin{ZX}
			\zxNone{}      &                			& \zxZ[a=pgp1]{\alpha_i}&
			\zxNone{}      &                			& \zxX[a=pgp1]{\pi}\\
			\zxNone{}      & 					& \zxX[a=pg1]{}	\ar[u, bend right] \ar[rrr, bend right] \ar[rrr, bend left] &
			\zxNone{}      & 					& \zxZ[a=pg1]{}	\ar[u, bend right]\\ 
			\zxNone{} \\
			\zxNone{} \rar & \zxX{} \ar[rrrruu] \rar 		& \zxNone{}\rar&
			\zxNone{} \rar & \zxZ{} \ar[lluu] \rar 			& \zxNone{}\\
			\zxNone{} \rar & \zxX{} \ar[rrrruuu] \rar 	 	& \zxNone{}\rar&
			\zxNone{} \rar & \zxZ{} \ar[lluuu] \rar		 	& \zxNone{}\\		
			\zxNone{}	& \vdots{}				&	&
			\zxNone{}	& \vdots{}				&	\\
			\zxNone{} \rar & \zxZ{}\ar[ruuuuu, bend right] \rar	& \zxNone{}\rar&
			\zxNone{} \rar 	& \zxX{}\ar[ruuuuu, bend right] \rar	& \zxNone{}\\
		\end{ZX}
		\overset{\eqref{eq:comp}}{=}\\
		&\begin{ZX}
			\zxNone{}      &                			& \zxZ[a=pgp1]{\alpha_i}&
			\zxNone{}      &                			& \zxX[a=pgp1]{\pi}\\
			\zxNone{}      & 					& \zxX[a=pg1]{}	\ar[u, bend right] &
			\zxNone{}      & 					& \zxZ[a=pg1]{}	\ar[u, bend right]\\ 
			\zxNone{} \\
			\zxNone{} \rar & \zxX{} \ar[rrrruu] \rar 		& \zxNone{}\rar&
			\zxNone{} \rar & \zxZ{} \ar[lluu] \rar 			& \zxNone{}\\
			\zxNone{} \rar & \zxX{} \ar[rrrruuu] \rar 	 	& \zxNone{}\rar&
			\zxNone{} \rar & \zxZ{} \ar[lluuu] \rar		 	& \zxNone{}\\		
			\zxNone{}	& \vdots{}				&	&
			\zxNone{}	& \vdots{}				&	\\
			\zxNone{} \rar & \zxZ{}\ar[ruuuuu, bend right] \rar	& \zxNone{}\rar&
			\zxNone{} \rar 	& \zxX{}\ar[ruuuuu, bend right] \rar	& \zxNone{}\\
		\end{ZX}
		= \dots =
		\begin{ZX}
			\zxNone{}      &                			& \zxZ[a=pgp1]{\alpha_i}&
			\zxNone{}      &                			& \zxX[a=pgp1]{\pi}\\
			\zxNone{}      & 					& \zxX[a=pg1]{}	\ar[u, bend right] \ar[rrr, bend left] &
			\zxNone{}      & 					& \zxZ[a=pg1]{}	\ar[u, bend right]\\ 
			\zxNone{} \\
			\zxNone{} \rar & \zxX{} \ar[rrrruu] \rar 		& \zxNone{}\rar&
			\zxNone{} \rar & \zxZ{} \ar[lluu] \rar 			& \zxNone{}\\
			\zxNone{} \rar & \zxX{} \ar[rrrruuu] \rar 	 	& \zxNone{}\rar&
			\zxNone{} \rar & \zxZ{} \ar[lluuu] \rar		 	& \zxNone{}\\		
			\zxNone{}	& \vdots{}				&	&
			\zxNone{}	& \vdots{}				&	\\
			\zxNone{} \rar & \zxX{}\ar[rrrruuuuu, bend right] \rar	& \zxNone{}\rar&
			\zxNone{} \rar 	& \zxZ{}\ar[lluuuuu] \rar	& \zxNone{}\\
		\end{ZX}
	\end{aligned}
\end{equation*}
Note that since, by definition, we have an unequal amount of legs, there will be one remaining connection due to the bi-algebra rule.
As a next step, we used the following process to remove the last remaining connection:
\begin{equation*}
    \begin{aligned}
		&\begin{ZX}[math baseline=t1]
			\zxNone{}      &                			& \zxZ[a=pgp1]{\alpha_i}&
			\zxNone{}      &                			& \zxX[a=pgp1]{\pi}\\
			\zxNone{}      & 					& \zxX[a=pg1]{}	\ar[u, bend right] \ar[rrr, bend left] &
			\zxNone{}      & 					& \zxZ[a=pg1]{}	\ar[u, bend right]\\ 
			\zxNone{} \\
			\zxNone{} \rar & \zxX{} \ar[rrrruu] \rar 		& \zxNone{}\rar&
			\zxNone{} \rar & \zxZ{} \ar[lluu] \rar 			& \zxNone{}\\
			\zxNone[a=t1]{} \rar & \zxX{} \ar[rrrruuu] \rar 	 	& \zxNone{}\rar&
			\zxNone{} \rar & \zxZ{} \ar[lluuu] \rar		 	& \zxNone{}\\		
			\zxNone{}	& \vdots{}				&	&
			\zxNone{}	& \vdots{}				&	\\
			\zxNone{} \rar & \zxX{}\ar[rrrruuuuu, bend right] \rar	& \zxNone{}\rar&
			\zxNone{} \rar 	& \zxZ{}\ar[lluuuuu] \rar	& \zxNone{}\\
		\end{ZX} \overset{\eqref{eq:pi_copy_split}}{=} 
		\begin{ZX}[math baseline=t1]
			\zxNone{}  	&\zxNone{}      	&                			& \zxNone{}			& \zxNone{}      	&                				& \zxZ{\alpha_i}		\\
			\zxNone{} 	&\zxNone{}      	& 					& \zxX{\pi} \ar[rrr, bend left] & \zxNone{}      	& 						& \zxX{}\ar[u, bend right]	\\ 
			\zxNone{} 	&\zxNone{}      	& \zxNone{}				& \zxNone{} 			& \zxNone{}      	& \zxNone{}					& \zxNone{}			\\ 
			\zxNone{}\rar 	&\zxX{\pi} \rar 	& \zxX{} \ar[ru, bend right] \rar 	& \zxNone{}\rar			& \zxNone{} \rar 	& \zxZ{} \ar[ruu, bend right] \rar 		& \zxNone{}			\\
			\zxNone{}\rar 	&\zxX{\pi} \rar 	& \zxX{} \ar[ru, bend right] \rar 	& \zxNone{}\rar			& \zxNone{} \rar 	& \zxZ[a=t1]{} \ar[ruuu, bend right] \rar	& \zxNone{}			\\		
			\zxNone{} 	&\zxNone{}		& \vdots{} 				& \zxNone{}			& \zxNone{}		& \vdots{}					&				\\
			\zxNone{}\rar 	&\zxX{\pi} \rar 	& \zxX{} \ar[ru, bend right] \rar	& \zxNone{}\rar			& \zxNone{} \rar 	& \zxZ{}\ar[ruuuuu, bend right] \rar		& \zxNone{}			\\
		\end{ZX} \overset{\eqref{eq:fusion}}{=} \\
		&\begin{ZX}[math baseline=t1]
			\zxNone{}  	&\zxNone{}      					&                			& \zxNone{}		&\zxNone{}      &                			& \zxZ[a=pgp1]{\alpha_i}\\
			\zxNone{}  	&\zxNone{}      					& 					& \zxNone{} 		&\zxNone{}      & 					& \zxX[a=pg1]{\pi}\ar[u, bend right]\\ 
			\zxNone{}  	&\zxNone{}      					& \zxNone{}				& \zxNone{} 		&\zxNone{}      & \zxNone{}				& \zxNone{}	\\ 
			\zxNone{}\rar  	&\zxX{\pi} \ar[ru, bend right] \rar 		& \zxNone{}  \rar 			& \zxNone{}\rar		&\zxNone{} \rar & \zxZ{} \ar[ruu, bend right] \rar 	& \zxNone{}\\
			\zxNone{}\rar  	&\zxX[a=t1]{\pi} \ar[ru, bend right] \rar 	& \zxNone{} \rar 	 		& \zxNone{}\rar		&\zxNone{} \rar & \zxZ{} \ar[ruuu, bend right] \rar	& \zxNone{}\\		
			\zxNone{}  	&\zxNone{}					& \vdots{}				&			&\zxNone{}	& \vdots{}				&	\\
			\zxNone{}\rar  	&\zxX{\pi} \ar[ru, bend right]\rar 		& \zxNone{} \rar			& \zxNone{}\rar		&\zxNone{} \rar & \zxZ{}\ar[ruuuuu, bend right] \rar	& \zxNone{}\\
		\end{ZX} \overset{\eqref{eq:minus_pi}}{=}
		\begin{ZX}[math baseline=t1]
			\zxNone{}  	&\zxNone{}     					&                	& \zxNone{} 	&\zxNone{}      &                			& \zxZ[a=pgp1]{-\alpha_i}\\
			\zxNone{}  	&\zxNone{}      				& 			& \zxNone{} 	&\zxNone{}      & 					& \zxX[a=pg1]{}	\ar[u, bend right]\\ 
			\zxNone{}  	&\zxNone{}      				& \zxNone{}		& \zxNone{} 	&\zxNone{}      & \zxNone{}				& \zxNone{}	\\ 
			\zxNone{}\rar  	&\zxX{\pi} \ar[ru, bend right] \rar 		& \zxNone{}  \rar 	& \zxNone{}\rar	&\zxNone{} \rar & \zxZ{} \ar[ruu, bend right] \rar 	& \zxNone{}\\
			\zxNone{}\rar  	&\zxX[a=t1]{\pi} \ar[ru, bend right] \rar 	& \zxNone{} \rar 	& \zxNone{}\rar	&\zxNone{} \rar & \zxZ{} \ar[ruuu, bend right] \rar	& \zxNone{}\\		
			\zxNone{}  	&\zxNone{}					& \vdots{}		&		&\zxNone{}	& \vdots{}				&	\\
			\zxNone{}\rar  	&\zxX{\pi} \ar[ru, bend right] \rar 		& \zxNone{} \rar	& \zxNone{}\rar	&\zxNone{} \rar & \zxZ{}\ar[ruuuuu, bend right] \rar	& \zxNone{}\\
		\end{ZX} \overset{\eqref{eq:pi_copy_split}}{=} \\&
		\begin{ZX}[math baseline=t1]
			\zxNone{}  	&\zxNone{}      				&                	& \zxX{\pi}			&\zxNone{}      	&                			& \zxZ[a=pgp1]{-\alpha_i}\\
			\zxNone{}  	&\zxNone{}      				& 			& \zxZ{}\ar[u, bend right] 	&\zxNone{}      	& 					& \zxX[a=pg1]{}	\ar[u, bend right]\\ 
			\zxNone{}  	&\zxNone{}      				& \zxNone{}		&  				&\zxNone{}      	& \zxNone{}				& \zxNone{}	\\ 
			\zxNone{}\rar  	&\zxX{} \ar[rruu, bend right] \rar 		& \zxNone{}  \rar 	& \zxNone{}\rar			&\zxNone{}\rar 		& \zxZ{} \ar[ruu, bend right] \rar 	& \zxNone{}\\
			\zxNone{}\rar  	&\zxX[a=t1]{} \ar[rruuu, bend right] \rar 	& \zxNone{} \rar 	& \zxNone{}\rar			&\zxNone{} \rar 	& \zxZ{} \ar[ruuu, bend right] \rar	& \zxNone{}\\		
			\zxNone{}  	&\zxNone{}					& \vdots{}		&				&\zxNone{}		& \vdots{}				&	\\
			\zxNone{}\rar  	&\zxX{} \ar[rruuuuu, bend right] \rar 		& \zxNone{} \rar	& \zxNone{}\rar			&\zxNone{} \rar 	& \zxZ{}\ar[ruuuuu, bend right] \rar	& \zxNone{}\\
		\end{ZX}
	\end{aligned}
\end{equation*}

\clearpage
\bibliographystyle{plain}
\bibliography{References.bib}

\begin{thebibliography}{10}

\bibitem{Akra1998}
Mohamad Akra and Louay Bazzi.
\newblock On the solution of linear recurrence equations.
\newblock {\em Computational Optimization and Applications}, 10:195--210, 1998.

\bibitem{Amy2019}
Matthew Amy and Vlad Gheorghiu.
\newblock staq -- {A} full-stack quantum processing toolkit.
\newblock {\em arXiv:1912.06070}, 2019.

\bibitem{Amy_2014}
Matthew Amy, Dmitri Maslov, and Michele Mosca.
\newblock Polynomial-time {T}-depth optimization of {C}lifford+{T} circuits via
  matroid partitioning.
\newblock {\em IEEE Transactions on Computer-Aided Design of Integrated
  Circuits and Systems}, 33(10):1476--1489, 2014.

\bibitem{Backens2014}
Miriam Backens.
\newblock The {ZX}-calculus is complete for stabilizer quantum mechanics.
\newblock {\em New J. Phys.}, 16(9):093021, 2014.

\bibitem{Backens2015}
Miriam Backens.
\newblock Making the stabilizer {ZX}-calculus complete for scalars.
\newblock In Chris Heunen, Peter Selinger, and Jamie Vicary, editors, {\em
  Proceedings of the 12th International Workshop on Quantum Physics and Logic},
  volume 195 of {\em Electronic Proceedings in Theoretical Computer Science},
  pages 17--32. Open Publishing Association, 2015.

\bibitem{Coecke2011}
Bob Coecke and Ross Duncan.
\newblock Interacting quantum observables: categorical algebra and
  diagrammatics.
\newblock {\em New J. Phys.}, 13(4):043016, 2011.

\bibitem{Coecke2017}
Bob Coecke and Aleks Kissinger.
\newblock {\em Picturing {Q}uantum {P}rocesses}.
\newblock Cambridge University Press, 2017.

\bibitem{Cowtan2019}
Alexander Cowtan, Silas Dilkes, Ross Duncan, Alexandre Krajenbrink, Will
  Simmons, and Seyon Sivarajah.
\newblock On the qubit routing problem.
\newblock 2019.

\bibitem{cowtan_et_al:LIPIcs:2019:10397}
Alexander Cowtan, Silas Dilkes, Ross Duncan, Alexandre Krajenbrink, Will
  Simmons, and Seyon Sivarajah.
\newblock On the qubit routing problem.
\newblock In Wim van Dam and Laura Mancinska, editors, {\em 14th Conference on
  the Theory of Quantum Computation, Communication and Cryptography (TQC
  2019)}, volume 135 of {\em Leibniz International Proceedings in Informatics
  (LIPIcs)}, pages 5:1--5:32, Dagstuhl, Germany, 2019. Schloss
  Dagstuhl--Leibniz-Zentrum fuer Informatik.

\bibitem{Cowtan2020}
Alexander Cowtan, Silas Dilkes, Ross Duncan, Will Simmons, and Seyon Sivarajah.
\newblock Phase gadget synthesis for shallow circuits.
\newblock {\em Electronic Proceedings in Theoretical Computer Science, EPTCS},
  318:213--228, 5 2020.

\bibitem{Duncan2019}
Ross Duncan, Aleks Kissinger, Simon Perdrix, and John van~de Wetering.
\newblock Graph-theoretic simplification of quantum circuits with the
  zx-calculus.
\newblock 2 2019.

\bibitem{Duncan2009}
Ross Duncan and Simon Perdrix.
\newblock Graph states and the necessity of {E}uler decomposition.
\newblock In Klaus Ambos-Spies, Benedikt L{\"o}we, and Wolfgang Merkle,
  editors, {\em Mathematical Theory and Computational Practice}, pages
  167--177, Berlin, Heidelberg, 2009. Springer Berlin Heidelberg.

\bibitem{PauliOpt}
Stefano Gogioso and Richie Yeung.
\newblock Annealing optimisation of mixed {ZX} phase circuits.
\newblock {\em arXiv:2206.11839}, 2022.

\bibitem{Jeandel2017}
Emmanuel Jeandel, Simon Perdrix, and Renaud Vilmart.
\newblock A complete axiomatisation of the {ZX}-calculus for {C}lifford+{T}
  quantum mechanics.
\newblock In {\em Proceedings of the 33rd Annual ACM/IEEE Symposium on Logic in
  Computer Science}, LICS '18, pages 559--568, 2018.

\bibitem{Jeandel2018}
Emmanuel Jeandel, Simon Perdrix, and Renaud Vilmart.
\newblock Diagrammatic reasoning beyond {C}lifford+{T} quantum mechanics.
\newblock In {\em Proceedings of the 33rd Annual ACM/IEEE Symposium on Logic in
  Computer Science}. {ACM}, 2018.

\bibitem{Kissinger2019}
Aleks Kissinger and Arianne~Meijer van~de Griend.
\newblock {CNOT} circuit extraction for topologically-constrained quantum
  memories.
\newblock {\em arXiv:1904.00633}, 4 2019.

\bibitem{Kissinger2021PYZX}
Aleks Kissinger and John van~de Wetering.
\newblock Simulating quantum circuits with zx-calculus reduced stabiliser
  decompositions.
\newblock 9 2021.

\bibitem{GriendPP}
Arianne Meijer-van~de Griend and Ross Duncan.
\newblock Architecture-aware synthesis of phase polynomials for {NISQ} devices.
\newblock 4 2020.

\bibitem{Nam2018}
Yunseong Nam, Neil~J. Ross, Yuan Su, Andrew~M. Childs, and Dmitri Maslov.
\newblock Automated optimization of large quantum circuits with continuous
  parameters.
\newblock {\em npj Quantum Inf.}, 4, 5 2018.

\bibitem{Nash_2020}
Beatrice Nash, Vlad Gheorghiu, and Michele Mosca.
\newblock Quantum circuit optimizations for {NISQ} architectures.
\newblock {\em Quantum Sci. Technol.}, 5(2):025010, mar 2020.

\bibitem{Sivarajah_2020}
Seyon Sivarajah, Silas Dilkes, Alexander Cowtan, Will Simmons, Alec Edgington,
  and Ross Duncan.
\newblock t$|$ket$\rangle$: a retargetable compiler for {NISQ} devices.
\newblock {\em Quantum Sci. Technol.}, 6(1):014003, 2020.

\bibitem{Wetering_ZXCalc}
John van~de Wetering.
\newblock {ZX}-calculus for the working quantum computer scientist.
\newblock {\em arXiv:2012.13966}, 12 2020.

\bibitem{Vilmart2019}
Renaud Vilmart.
\newblock {\em {ZX}-calculi for quantum computing and their completeness}.
\newblock PhD thesis, 09 2019.

\bibitem{Winderl2022ZXPolynomial}
David Winderl.
\newblock {ZX} polynomial synthesis.
\newblock {\em Preprint}, 2022.

\bibitem{Yeung2020}
Richie Yeung.
\newblock Diagrammatic design and study of ans\"atze for quantum machine
  learning.
\newblock {\em arXiv:2011.11073}, 2020.

\bibitem{osti_1785933}
Ed~Younis, Costin~C. Iancu, Wim Lavrijsen, Marc Davis, Ethan Smith, and USDOE.
\newblock Berkeley quantum synthesis toolkit ({BQSKit}) v1, 2021.

\end{thebibliography}
\end{document}